\documentclass[aps,pra,reprint]{revtex4-1}

\usepackage{xcolor}
\usepackage{tikz}
\usepackage{amsmath}
\usepackage{amssymb,amsmath}
\usepackage{graphicx}
\usepackage{upgreek}
\usepackage[T1]{fontenc}
\usepackage[utf8]{inputenc}
\usepackage{graphicx}
\usepackage{bm}

\begin{document}

\title{Quantum metamaterials with magnetic response at optical frequencies}

\author{Rasoul Alaee,{$^{1,2,\ast}$} Burak Gurlek,{$^{1}$} Mohammad Albooyeh,{$^{3}$} Diego Mart\'{i}n-Cano,{$^{1}$} and Vahid Sandoghdar{$^{1,\ast}$}}
\address{$^1$Max Planck Institute for the Science of Light, Erlangen 91058, Germany}
\address{$^2$Department of Physics, University of Ottawa, Ottawa Q1N 6N5, Canada}
\address{$^3$Department of Electrical Engineering and Computer Science, University of California, Irvine, CA 92617, USA}
\address{$^{\ast}$\rm Emails: rasoul.alaee@mpl.mpg.de and vahid.sandoghdar@mpl.mpg.de}


\begin{abstract}
We propose novel quantum antennas and metamaterials with strong \textit{magnetic} response at optical frequencies. Our design is based on the arrangement of natural atoms with only \textit{electric} dipole transition moments at distances smaller than a wavelength of light but much larger than their physical size. In particular, we show that an atomic dimer can serve as a magnetic antenna at its antisymmetric mode to enhance the decay rate of a magnetic transition in its vicinity by several orders of magnitude. Furthermore, we study metasurfaces composed of atomic bilayers with and without cavities and show that they can fully reflect the electric and magnetic fields of light, thus, forming nearly perfect electric/magnetic mirrors. The proposed quantum metamaterials can be fabricated with available  state-of-the-art technologies and promise several applications both in classical optics and quantum engineering.
\end{abstract}



\maketitle	
Most natural materials interact weakly with the magnetic field of light at optical frequencies~\cite{LandauLifshitz}. In fact, the magnetic interaction energy $\bm{-{\mu}\cdot\mathbf{B}}$ is typically about two orders of magnitude (i.e. order of fine-structure constant) smaller than its electric counterpart $\mathbf{-p\cdot\mathbf{E}}$, whereby $\mu\approx\mu_B$ and $p\approx ea_{0}$ represent the magnitude of the electric dipole moments, and $e$, $a_{0}$, $\mu_B$, denote the elementary charge, Bohr radius, and Bohr magneton, respectively~\cite{Jackson1999}. However, two decades of progress in nano-optics has brought about novel electromagnetic properties that are not available in natural materials. In particular, ``metamaterials'' created through synthetic arrangement of subwavelength antennas~\cite{engheta:2006,Soukoulis:2011,Yu2014,Kuznetsov2016} can now generate magnetic functionalities at high frequencies ~\cite{Shalaev:Book2008}. Unfortunately, material absorption and limits in nanofabrication have hampered reaching a high performance in the optical regime.
 
Considering that natural atoms act as the smallest and most fundamental optical antennas~\cite{,Zumofen:2008,Sandoghdar:Book2013}, one can also envision the construction of ``quantum'' metamaterials by synthetically arranging natural atoms or molecules at distances smaller than an optical wavelength but much larger than the characteristic length of electronic orbitals. Indeed, a number of such proposals have emerged over the past few years~\cite{meir2013Lambshift,Bettles2016,shahmoon2017,Zhou2017,mkhitaryan2018,wild2018quantum,Manzoni2018,Genes2019,plankensteiner2019enhanced,grankin2018,liberal2018,Zoller2019,bettles2019quantum,Rui:2020}, but these have only considered metamaterials with electric response. In this Letter, we show that a strong magnetic functionality can be obtained from conventional atoms at optical frequencies. In particular, we propose novel quantum antennas that can  enhance the decay rate of a magnetic emitter (i.e. an emitter with magnetic dipole transitions) in their vicinity by several orders of magnitude. We demonstrate that a metasurface composed of the proposed antennas can act as nearly perfect electric and magnetic mirrors and can, moreover, strongly couple to a cavity mode independent of its position.

\begin{figure}
\begin{centering}
\includegraphics[width=0.5\textwidth]{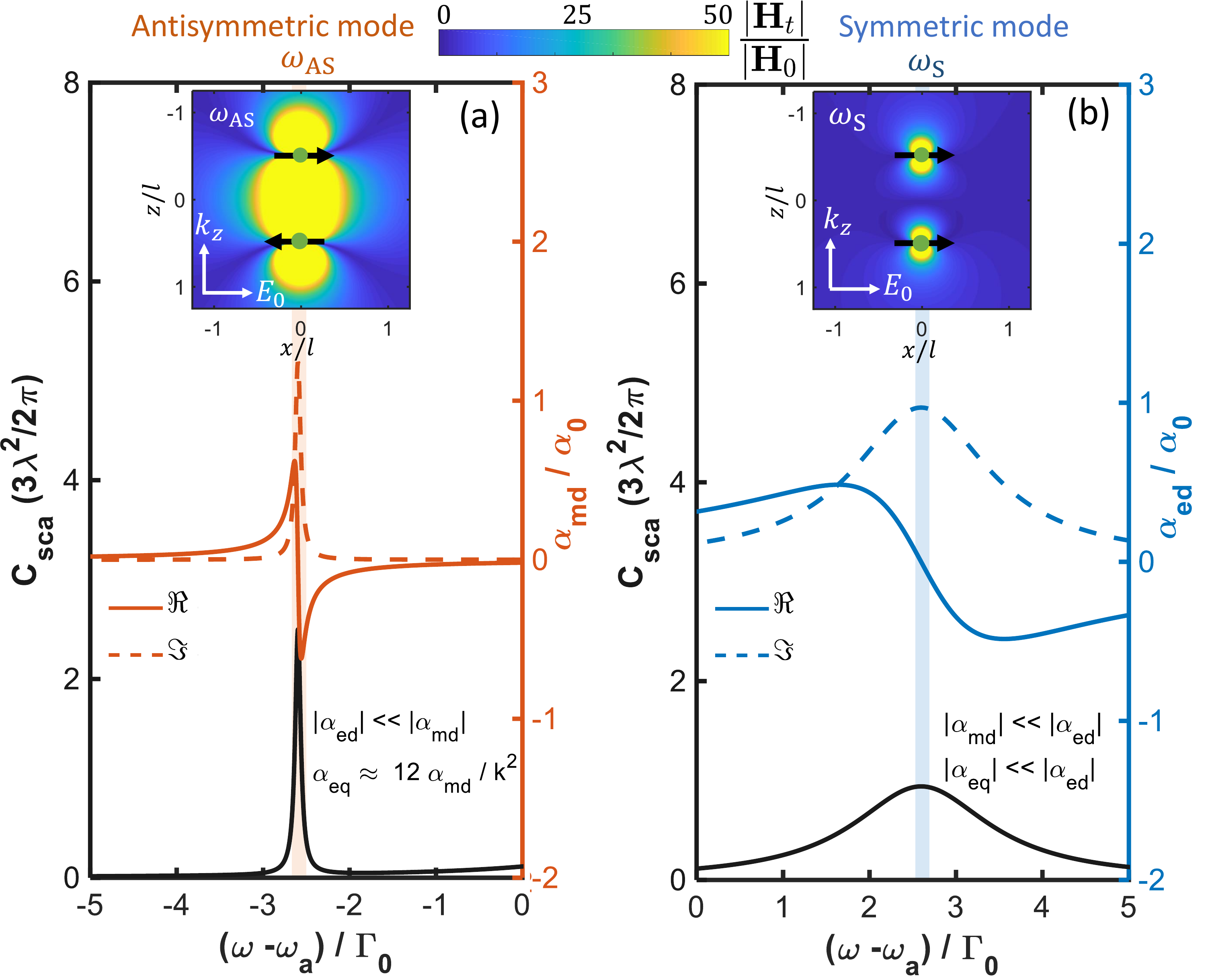}
\par\end{centering}
\caption{\textit{Atomic dimer antenna:} Optical response of two atoms placed at $z=\pm\frac{l}{2}$ (see insets) as a function of detuning at the antisymmetric (a) and symmetric (b) modes of the composite system, respectively. Left vertical axes (black): total scattering cross sections normalized to the free-space value $\frac{3}{2\pi}\lambda^{2}$ for a two-level atom on resonance. Right vertical axes (orange and blue): The real (solid curves) and imaginary (dashed curves) parts of the induced effective polarizabilities calculated using Eq.~\ref{eq:ME} and for $l=0.1\lambda_a$. Insets display normalized total magnetic field distribution for each case. The antisymmetric mode exhibits both magnetic and electric quadrupole response which lead to a larger total cross section than the symmetric mode~\cite{Rahimzadegan:17}.}\label{fig:AtomicDimer}
\end{figure}

\textit{Atomic dimer antenna.—}Let us first consider an atomic dimer consisting of two identical atoms with electric dipole transition moments placed at $\mathbf{r}_{u/d}=(0,0,\pm l/2)$~(see the inset in Fig.~\ref{fig:AtomicDimer} (a,b); u,d stand for up and down). We assume that the atomic response is isotropic and linear, i.e. we consider the weak-excitation limit. The electric polarizability of each atom amounts to $\alpha\left(\omega\right)=\frac{-\frac{\Gamma_{0}}{2}\alpha_{0}}{\delta+i\frac{\Gamma_{0}}{2}}$, where $\Gamma_{0}$ is the radiative linewidth of the atomic transition at frequency $\omega_{a}$ while $\delta=\omega-\omega_{a}\ll\omega_{a}$ represents the frequency detuning between the illumination and the atom, $\alpha_{0}=6\pi/k^{3}$ and $k$ is the wavenumber~\cite{lambropoulos2007}. We note that our discussion can be readily generalized to any other quantum emitter such as molecules, color centers, quantum dots or ions with dipolar transitions.

The atomic dimer antenna is illuminated by an $x$-polarized plane wave $\mathbf{E}_{\mathrm{inc}}=E_{0}e^{ikz}\mathbf{e}_{x}$ propagating in the $z$ direction, where $\mathbf{e}_{x}$ is the unit vector in the $x$ direction, $E_{0}$ is the electric field amplitude, and $k$ is the wavenumber in free space. The induced volume current density for the dimer antenna can be written as (see the Supplementary Materials (SM) for a detailed derivation)
\begin{equation}
\mathbf{J}\left(\mathbf{r},\omega\right) =  -i\omega\epsilon_{0}E_{0}\left[\alpha_{d}\delta\left(\mathbf{r}-\mathbf{r}_{d}\right)+\alpha_{u}\delta\left(\mathbf{r}-\mathbf{r}_{u}\right)\right]\mathbf{e}_{x},\label{eq:Jind}
\end{equation}
where $\delta\left(\mathbf{r}-\mathbf{r}_{u/d}\right)$ is the Dirac delta function and $\epsilon_{0}$ denotes the free-space permittivity. The quantities $\alpha_{u}$ and $\alpha_{d}$ are the effective electric polarizabilities of the upper and lower atoms according to $\alpha_{u/d}= \alpha\left[\frac{\mathrm{cos}\left(kl/2\right)}{D_{-}}\pm i\frac{\mathrm{sin}\left(kl/2\right)}{D_{+}}\right]$. Here, $D_{\pm}\equiv1\pm\epsilon_{0}\alpha G_{EE}^{xx}\left(\mathbf{r}_{u},\mathbf{r}_{d}\right)$ with $G_{EE}^{xx}\left(\mathbf{r}_{d},\mathbf{r}_{u}\right)$ signifying the scalar Green\textquoteright s function of the Helmholtz equation in free space. Using a multipole expansion of the induced current~\cite{Alaee:2018} and Eq.~(\ref{eq:Jind}), we obtain the induced electric and magnetic
dipole and quadrupole polarizabilities of the dimer antenna at $\mathbf{r}=0$~(see SM):
\begin{eqnarray}
\alpha_{\rm ed}	& = &	\frac{2\alpha}{D_{-}}\left[j_{0}\left(kl/2\right)-\frac{j_{2}\left(kl/2\right)}{2}\right]\mathrm{cos}\left(kl/2\right),\nonumber \\
\alpha_{\rm md}	& = &\frac{3\alpha}{D_{+}}j_{1}\left(kl/2\right)\mathrm{sin}\left(kl/2\right),\nonumber
\end{eqnarray}
\begin{eqnarray}
\alpha_{\rm eq}	& = &\frac{12}{k^{2}}\frac{\alpha}{D_{+}}\left[3j_{1}\left(kl/2\right)-2j_{3}\left(kl/2\right)\right]\mathrm{sin}\left(kl/2\right),\nonumber \\
\alpha_{\rm mq}	& = &-\frac{60}{k^{2}}\frac{\alpha}{D_{-}}j_{2}\left(kl/2\right)\mathrm{cos}\left(kl/2\right),\label{eq:ME}
\end{eqnarray}
where $j_{n}\left(r\right)$ is the spherical Bessel function and $\alpha_{\rm ed}$, $\alpha_{\rm md}$, $\alpha_{\rm eq}$, and $\alpha_{\rm mq}$ represent the electric dipole, magnetic dipole, electric quadrupole and magnetic quadrupole polarizabilities, respectively. For small separations ($l\ll \lambda$), the higher order spherical Bessel functions are negligible, i.e. $j_{2}\left(kl/2\right)\approx0$ and $j_{3}\left(kl/2\right)\approx0$. Thus, $\alpha_{\rm md}\approx k^2/12\alpha_{\rm eq}$ and the magnetic quadrupole polarizability can be neglected, i.e. $\alpha_{\rm mq}\approx0$. Once the induced dipole and quadrupole polarizabilities are obtained, the total scattering cross section ($C_{\mathrm{sca}}$) of the atomic dimer can also be calculated (see SM):
\begin{equation}
\label{eq:C_sca}
C_{\mathrm{sca}}=\frac{k^{4}}{6\pi}\left(\left|\alpha_{{\rm ed}}\right|^{2}+\left|\alpha_{{\rm md}}\right|^{2}+\frac{3}{5}\left|\frac{k^{2}}{12}\alpha_{{\rm eq}}\right|^{2}\right).
\end{equation}

Near-field coupling of the electric dipole transitions of two individual emitters has been explored in various systems~\cite{DeVoe:1996,Hettich:2002} and is known to lead to symmetric (superradiant) and antisymmetric (subradiant) states. The black curve in Fig.~\ref{fig:AtomicDimer}(a) shows $C_{\mathrm{sca}}$ for the subradiant state as a function of the frequency detuning for two atoms separated by $l=\lambda_{a}/10$. In this case, the electric response of the dimer antenna becomes negligible, but it exhibits both magnetic dipolar (see right vertical axis) and electric quadrupolar responses with $\alpha_{\rm eq}\approx\frac{12}{k^{2}}\alpha_{\rm md}$ (see SM). The inset in Fig.~\ref{fig:AtomicDimer}(a) illustrates the magnetic field distribution for this antisymmetric mode under plane wave illumination, where a strong magnetic field testifies to an optically induced magnetic response. The left vertical axis in Fig.~\ref{fig:AtomicDimer}(b) shows $C_{\mathrm{sca}}$ as a function of the frequency detuning for the symmetric mode, where the two atoms oscillate in phase. The right vertical axis plots the electric dipolar response of the antenna structure, while the inset shows that the magnetic response is negligible in this scenario. 

\begin{figure}
\begin{centering}
\includegraphics[width=0.48\textwidth]{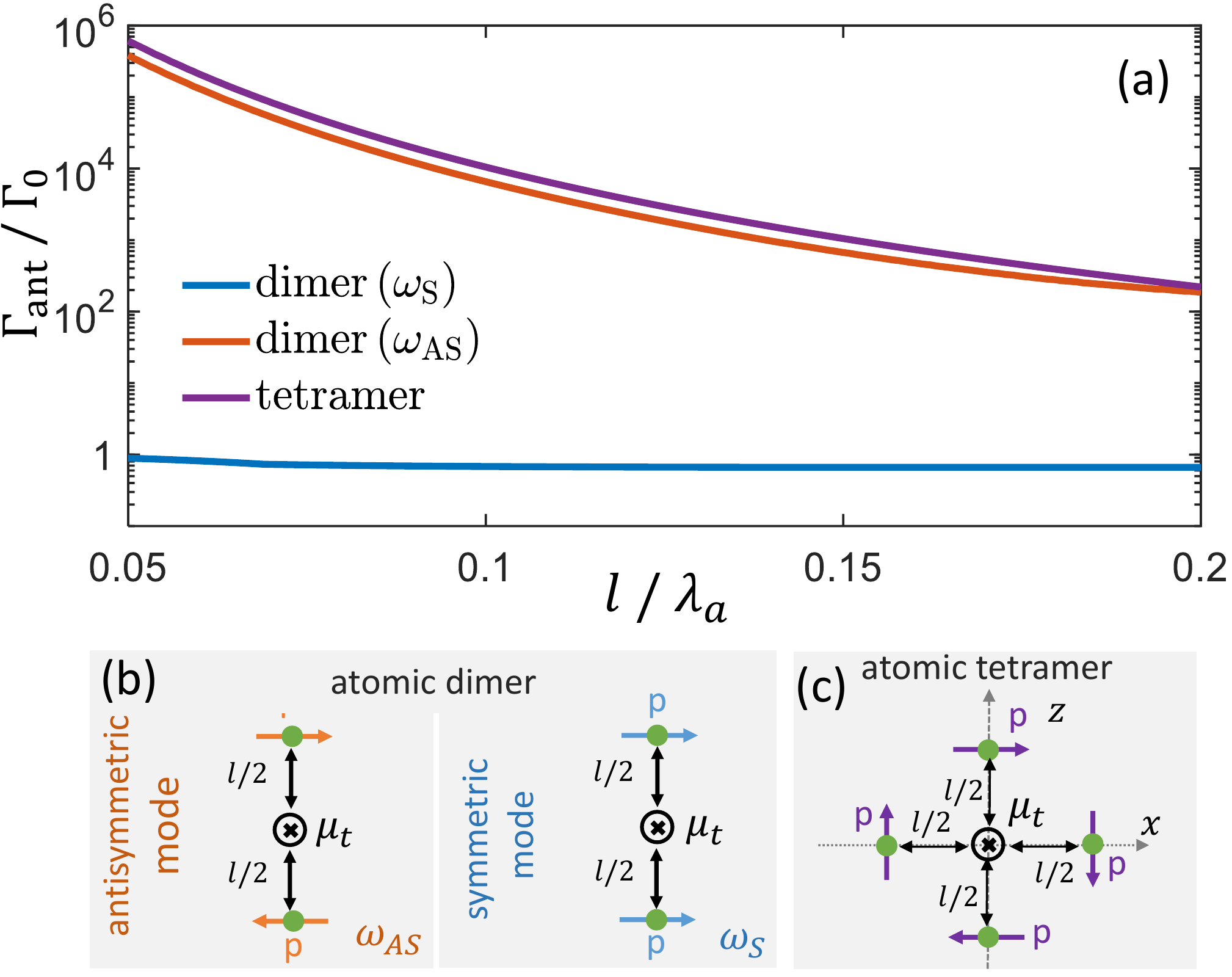}
\par\end{centering}

\caption{\textit{Enhancing the decay rate of a magnetic transition}: (a) Enhanced decay rate of a magnetic dipole emitter placed in the middle of an atomic dimer antenna for the symmetric (blue), antisymmetric (red) modes and an atomic tetramer (purple) as a function of the antenna length $l$. The tetramer antenna is composed of four identical atoms placed at  $\mathbf{r}_{1,2}=\mp l/2\,\mathbf{e}_{z}, \mathbf{r}_{3,4}=\pm l/2 \,\mathbf{e}_{x}$. (b,c) Schematics of an emitter with magnetic dipole moment $\mu_{t}$ placed in the middle of an atomic dimer (b) and tetramer (c). \label{fig:LDOS}}
\end{figure}

\textit{Enhancing magnetic transitions—} The strong magnetic field generated in the atomic dimer ~(see Fig.~\ref{fig:AtomicDimer}(a)) prompts us to inquire whether it can act as a magnetic antenna to enhance the decay of a test magnetic dipole moment $\mu_{t}$ placed at the origin. Using the normalized local density of states (NLDOS) of the system (see SM), one can arrive at the antenna-modified decay rate $\Gamma_{\mathrm{\rm ant}}$ given by
\begin{eqnarray}
\frac{\Gamma_{\mathrm{\rm ant}}}{\Gamma_{0}}=1-\epsilon_{0}^{2}\alpha_{0}\alpha\mathrm{Im}\left[\frac{g_{\rm EM}^{2}\left(\mathbf{r}_{0},\mathbf{r}_{u}\right)+g_{\rm EM}^{2}\left(\mathbf{r}_{0},\mathbf{r}_{d}\right)}{1+\epsilon_{0}\alpha G_{\rm EE}^{xx}\left(\mathbf{r}_{u},\mathbf{r}_{d}\right)}\right],
\label{eq:Decay}
\end{eqnarray}
where $g_{\rm EM}\left(\mathbf{r},\mathbf{r}^{\prime}\right)=-\frac{3}{2\epsilon_{0}\alpha_{0}}e^{i\zeta}\left(\frac{1}{\zeta}+\frac{i}{\zeta^{2}}\right)$ is the scalar electro-magnetic Green\textquoteright s function in free space and $\zeta=k\left|\mathbf{r}-\mathbf{r}^{\prime}\right|$~(see SM). Figure~\ref{fig:LDOS}(a) plots the calculated magnetic decay rate enhancement for both symmetric and antisymmetric modes. We find that the decay rate can be enhanced by \textit{five} orders of magnitude at $l\approx\lambda_{a}/20$ for the antisymmetric mode. However, for the symmetric mode the decay rate is even slightly decreased below its unperturbed value (i.e., $\Gamma_{\mathrm{ant}}<\Gamma_{0}$) because of the weak magnetic response of this mode~(see the inset in Fig.~\ref{fig:AtomicDimer}(b)). 

To achieve even larger enhancements, one can devise an atomic tetramer antenna, consisting of four identical atoms with electric polarizability $\alpha$ (see Fig.~\ref{fig:LDOS}(c)). Figure~\ref{fig:LDOS}(a) shows the enhancement of the magnetic transition rate for this case (see SM). We note that fabrication of quantum metamaterials in these configurations is readily within reach since the distances involved are well beyond atomic and molecular spacings in natural substances (e.g.~$\lambda_{a}/20$ corresponds to several tens of nanometers). A particularly interesting class of materials for these applications are rare earth ions with weak magnetic dipole transition \cite{Karaveli:11, Kasperczyk2015}. New efforts on the implantation of ions using ion traps or other bombardment strategies \cite{GrootBerning:2019, Luhmann:2018} allow precise doping of various host materials.
\begin{figure}
\begin{centering}
\includegraphics[width=0.5\textwidth]{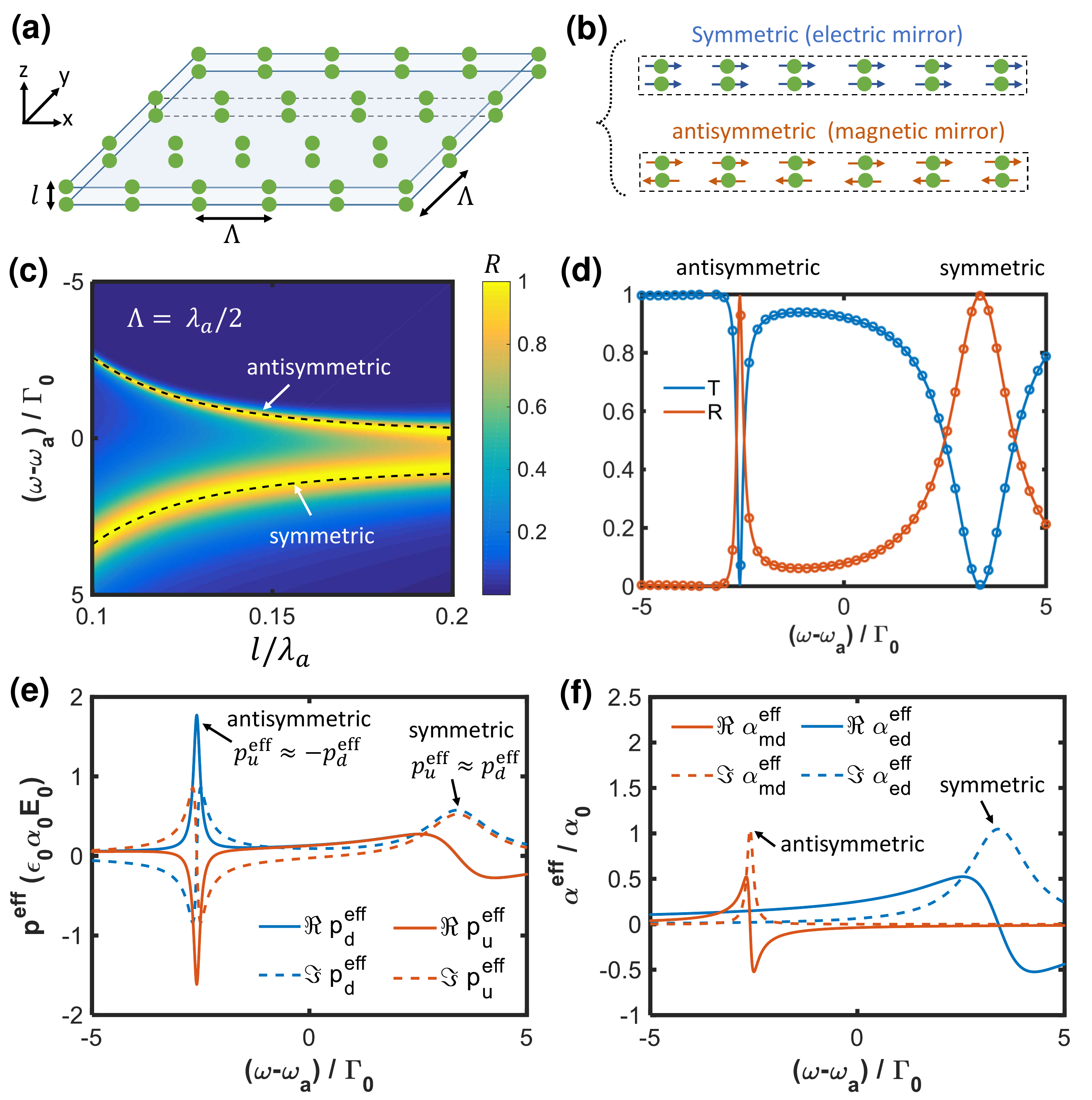}
\par\end{centering}

\caption{Atomic bilayer metasurface (ABM): (a) Schematics of an ABM. (b) The symmetric and antisymmetric modes corresponding to the electric and magnetic mirrors, respectively. (c) Intensity reflection coefficient $R$ as a function of layer separation $l$ and frequency detuning $\omega-\omega_a$ for an ABM with periodicity $\Lambda_{x}=\Lambda_{y}=\lambda_{a}/2$. (d) Intensity transmission and reflection coefficients corresponding to a cut through (c) at $l=0.1\lambda_a$. Solid curves: analytical results for an infinite array with plane wave illumination. Symbols: Numerical calculations for a finite array with $15\times15\times2$ atoms illuminated by a Gaussian beam. (e) The real and imaginary parts of the effective induced dipole moments of different layers. (f) The real and imaginary parts of the effective electric and magnetic polarizabilities. 
\label{fig:Atomic_Metasurface}} 
\end{figure}


\textit{Electric and magnetic mirrors based on atomic bilayer metasurfaces.—} It has been shown that optimal optical coupling to a two-level atom requires mode matching between the incident light and that of the field radiated by the atom~\cite{Zumofen:2008}. It is, thus, found that a dipolar wave can be perfectly reflected by a single two-level atom with a dipolar transition. Similarly, it has been shown, both theoretically and experimentally, that a planar two-dimensional array of atoms acts as a nearly perfect electric mirror for a plane-wave illumination~\cite{Bettles2016,shahmoon2017,Rui:2020}. Now, we show that a periodic planar arrangement of our dimer antennas, which we call atomic bilayer metasurfaces (ABM), can act as both electric and magnetic mirrors (see Fig.~\ref{fig:Atomic_Metasurface}(a,b)). 

Let us illuminate an ABM by an $x$-polarized plane wave propagating in $z$-direction, $\mathbf{E}_{\mathrm{inc}}=E_{0}e^{ikz}\mathbf{e}_{x}$. The field
reflection and transmission coefficients are given by (see SM)
\begin{eqnarray}
r & = & \frac{ik}{2\Lambda^{2}\epsilon_{0}E_{0}}\left(p_{d}^{\mathrm{eff}}e^{-ikl/2}+p_{u}^{\mathrm{eff}}e^{ikl/2}\right),\nonumber \\
t & = & 1+\frac{ik}{2\Lambda^{2}\epsilon_{0}E_{0}}\left(p_{d}^{\mathrm{eff}}e^{ikl/2}+p_{u}^{\mathrm{eff}}e^{-ikl/2}\right),\label{eq:TR_2Layers}
\end{eqnarray}
where $p_{u}^{\mathrm{eff}}$ and $p_{d}^{\mathrm{eff}}$ are the
effective electric dipole moments of the upper and lower layers, respectively and can be calculated
as
\begin{eqnarray}
\left[\begin{array}{c}
p_{d}^{\mathrm{eff}}\\
p_{u}^{\mathrm{eff}}
\end{array}\right] & = & \left[\begin{array}{cc}
\frac{1}{\epsilon_0\alpha}-C_{dd} & -C_{du}\\
-C_{ud} & \frac{1}{\epsilon_0\alpha}-C_{uu}
\end{array}\right]^{-1}\left[\begin{array}{c}
E_{i}\left(\mathbf{r}_{d}\right)\\
E_{i}\left(\mathbf{r}_{u}\right)
\end{array}\right]\label{eq:Induced_ED_eff}
\end{eqnarray}
with 
\begin{eqnarray}
C_{dd}=\sum_{n,\,n\neq0}G_{EE}^{xx}\left(\mathbf{r}_{d,0},\mathbf{r}_{d,n}\right) \\ C_{du}=\sum_{n}G_{EE}^{xx}\left(\mathbf{r}_{d,0},\mathbf{r}_{u,n}\right)\,,\label{eq:IC}
\end{eqnarray}
denoting the interaction constants between atoms in the upper and lower layers, respectively. For identical atoms, the interaction constants are symmetric, i.e. $C_{ud}=C_{du}$ and $C_{uu}=C_{dd}$.

For lossless atoms (i.e. no nonradiative loss), the imaginary part of the interaction constants can be calculated exactly by using the conservation of energy principle~[see SM],
\begin{eqnarray}
\mathrm{Im}\left[C_{dd}\right] = \frac{k}{2\Lambda^{2}\epsilon_{0}}-\frac{1}{\epsilon_0\alpha_{0}}\\ \, \mathrm{Im}\left[C_{du}\right] = \frac{k}{2\Lambda^{2}\epsilon_{0}}\,\mathrm{cos}\left(kl\right),
\end{eqnarray}
while their real parts are calculated numerically.
Figure~\ref{fig:Atomic_Metasurface}(c) shows the reflectivity of an ABM as a function of the frequency detuning and the distance between the two layers. It can be seen that the array fully reflects the impinging light at both asymmetric and antisymmetric resonance frequencies. In Fig.~\ref{fig:Atomic_Metasurface}(d), we plot the reflection and transmission of a plane wave incident on an infinite array for $l=\lambda_{a}/10$ calculated using Eq.~\ref{eq:TR_2Layers} (solid lines). The symbols in Fig.~\ref{fig:Atomic_Metasurface}(d) present the results obtained for a finite array of $15\times15\times2$ atoms and a Gaussian beam illumination (a possible experimental situation) by integrating the Poynting vector for the scattered and incident beam [see SM]. We find that the results for the two illuminations (i.e. Gaussian and plane wave) agree very well. 

We now provide more insight into the working of the ABM in Fig.~\ref{fig:Atomic_Metasurface}(e,f). As displayed in Fig.~\ref{fig:Atomic_Metasurface}(e), the effective induced dipole moments of the upper and lower layers are out of phase at the antisymmetric mode, i.e. $p_{u}^{\mathrm{eff}}\approx-p_{d}^{\mathrm{eff}}\approx1.7 \epsilon_0\alpha_0E_0$. Figure~\ref{fig:Atomic_Metasurface}(f) displays the effective electric and magnetic polarizabilities of the atomic metasurface, which can be calculated by using
\begin{equation}
\alpha_{\rm ed}^{\mathrm{eff}}=\frac{p_{d}^{\mathrm{eff}}+p_{u}^{\mathrm{eff}}}{\epsilon_0E_{0}}\mathrm{cos}\left(kl/2\right),
\alpha_{\rm md}^{\mathrm{eff}}=i\frac{p_{d}^{\mathrm{eff}}-p_{u}^{\mathrm{eff}}}{\epsilon_0E_{0}}\mathrm{sin}\left(kl/2\right).\nonumber\\
\end{equation}
It follows that the the antisymmetric resonance of the ABM supports an effective magnetic response $\alpha_{\rm md}^{\mathrm{eff}}\approx i\alpha_0$, while $\alpha_{\rm ed}^{\mathrm{eff}}\approx0$. Together with a reflectivity $r\approx1$, this implies that the bilayer metasurface acts as a nearly perfect \textit{atomic magnetic mirror} at its antisymmetric resonance. At the symmetric mode, however, the atoms in the upper and lower layers are in phase such that $p_{u}^{\mathrm{eff}}\approx p_{d}^{\mathrm{eff}}\approx0.5i\epsilon_0\alpha_0E_0$, leading to an effective electric response, i.e. $\alpha_{\rm ed}^{\mathrm{eff}}\approx i\alpha_0$, but
$\alpha_{\rm md}^{\mathrm{eff}}\approx0$ (see Fig.~\ref{fig:Atomic_Metasurface}(f)). Therefore, the array acts as a nearly perfect \textit{atomic electric mirror} with $r\approx-1$~(see SM for a Gaussian beam excitation).

\begin{figure}
\begin{centering}
\includegraphics[width=0.48\textwidth]{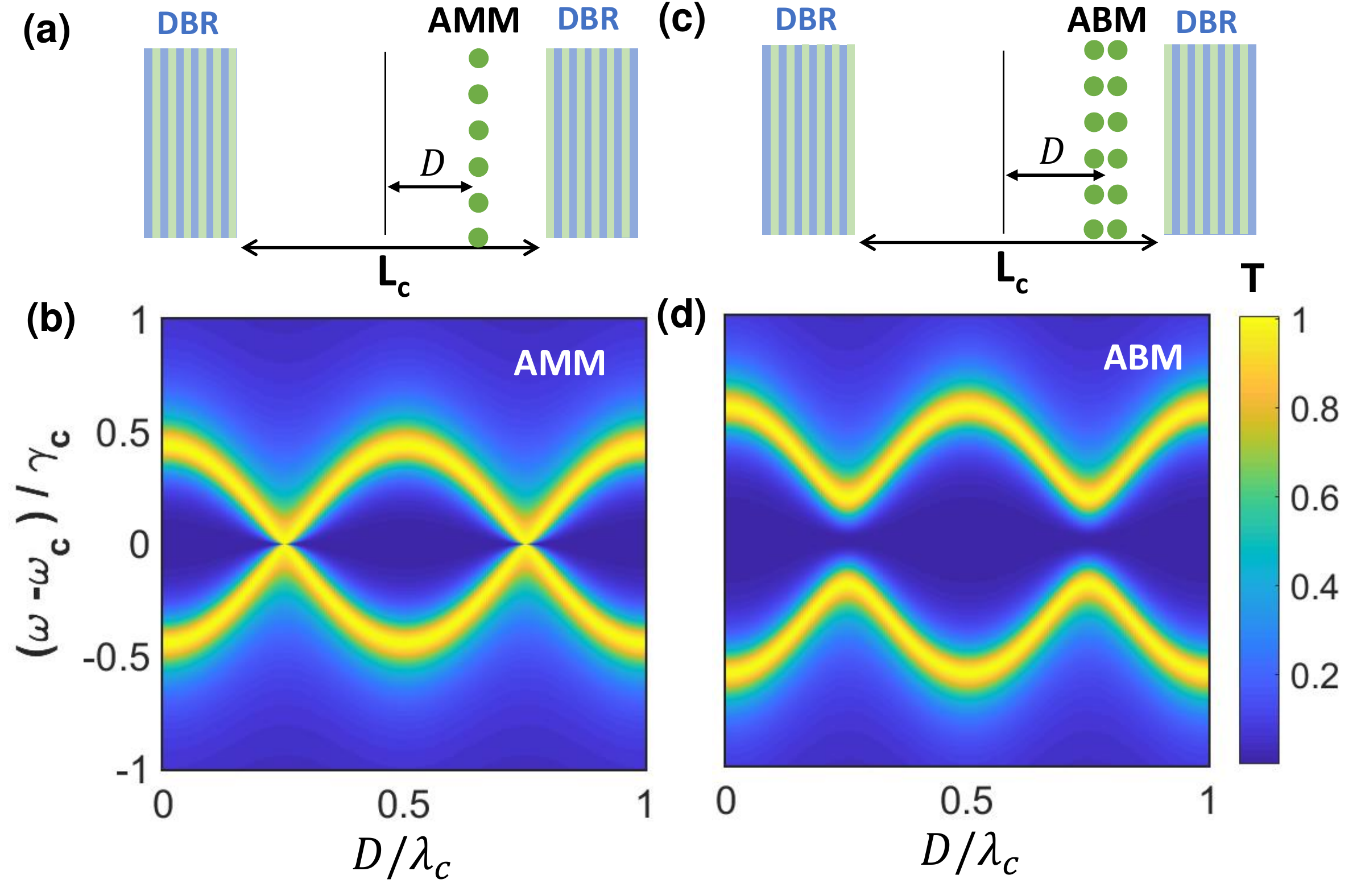}
\par\end{centering}

\caption{\textit{Atomic metasurface inside a planar cavity}: (a) Schematics of a cavity and an atomic monlayer metasurface (AMM). (b) Transmission of the cavity loaded with the AMM as a function of its position inside the cavity ($D$) and detuning. (c,d) Same as (a,b) but for an atomic bilayer metasurface (ABM). The quality factor of the planar cavity is taken to be $Q\approx8.7\times10^{3}$, $\gamma_{c}\approx10^{3}\Gamma_{0}$ and $L_{c}=3\lambda_a$. \label{fig:Cavity} }
\end{figure}


\textit{Atomic monolayer/bilayer metasurfaces in a cavity.—}Optical cavities are commonly used to enhance the interaction of light with matter. The maximum interaction occurs when an atom with an electric (magnetic) transition dipole moment is placed at the maximum electric (magnetic) field. In other words, the interaction strongly depends on the position of the atom inside the optical cavity. The proposed atomic bilayer metasurface can overcome this problem due to the combination of its strong electric and magnetic response to light.

Figure~\ref{fig:Cavity} depicts a planar cavity consisting of two mirrors separated by $L_{c}$, whereby the transmission of the bare cavity is assumed to reach unity at $\omega=\omega_{c}$, i.e., there are no absorption or scattering losses in the cavity. Now, we investigate the interaction of the cavity with an atomic metasurface placed at a distance $D$ from its center for i) an atomic monolayer metasurface (AMM), and ii) an ABM~(see Fig.~\ref{fig:Cavity}(a,c)). The cavity linewidth $\gamma_c$ is assumed to be much larger than that of the AMM resonance ($\gamma_{c}\approx10^{3}\Gamma_0$). The calculated transmission is plotted in Fig.~\ref{fig:Cavity}(b) for the AMM as a function of the frequency detuning and metasurface position $D$. It is seen that one reaches strong coupling at $D=0$, but the resonance splitting decreases with increasing $D$. The transmission changes periodically by varying the position $D$ of the metasurface. At $D=\lambda_{c}/4$ there is no longer a splitting due to a vanishing electric field. At $\omega=\omega_c$, transmission is zero~(see SM, section IV) because the first cavity mirror and the AMM (which is a nearly perfect electric mirror at the cavity frequency, i.e. $r_{\rm AMM}=-1$) form a new cavity. 

For an ABM, however, strong coupling can be maintained at all positions inside the cavity~(Fig.~\ref{fig:Cavity}(d)). While at $D=0$, the cavity only interacts with the symmetric mode due to the maximum electric field inside cavity, at $D=\lambda_{c}/4$ the cavity only couples to the antisymmetric mode. At intermediate positions, where $0<D<\lambda_{c}/4$, the cavity couples to both symmetric and antisymmetric modes.

In conclusion, we have demonstrated that synthetic arrangements of natural atoms with only electric dipole transitions can support both electric and magnetic responses at optical frequencies. Since the interatomic distances required for our proposed designs are well above ten nanometers, our proposed quantum metasurfaces can be experimentally realized in both the gas and solid phases using a range of available methods in cold atom manipulation or implantation strategies. In particular, our proposal lends itself to applications based on natural species with weak magnetic dipole transitions, e.g., rare earth ions, (See Refs~\cite{Dodson:2012,van2018recent}). For instance, $\mathrm{Eu^{3+}}$ with a magnetic transition ($^{5}D_{0}\rightarrow^{7}F_{1}$) at 584\,nm and a linewidth of about 15\,Hz (see Table III Ref~\cite{Dodson:2012}), is a suitable candidate for enhancement by an arrangement of different atoms such as Fe, Ar, Kr, N, Na with electric dipole transition at the same wavelength (see Ref~\cite{van2018recent}, J = 0 $\rightarrow$ J = 1 with an E1 transition). The predicted transition rate enhancements reaching $10^5$ would, thus, yield magnetic transitions with natural linewidths fully comparable to that of common electric dipole transitions. Such novel materials hold promise for the development of a range of technological applications and fundamental studies in quantum engineering and physics. 

\textbf{Acknowledgments.—} This work was supported by the Max Planck Society. R.A. also acknowledges financial support provided by the Alexander von Humboldt Foundation. The authors warmly thank Claudiu Genes for helpful discussions. 

\appendix

\begin{widetext}
\section{Atomic dimer}
\subsection{Induced multipole moments}
In this section, we derive induced multipole moments of an atomic
dimer. Let us consider an atomic dimer consisting of two identical
atoms placed at $\mathbf{r}_{u/d}=[0,0,\pm l/2]$ and illuminated
by an $x$-polarized plane wave propagating in the $z$ direction,
i.e. $\mathbf{E}_{\mathrm{inc}}=E_{0}e^{ikz}\mathbf{e}_{x}$ {[}see
Fig.~\ref{fig:Geometry_dimer} (a)-(b){]}. We assume $e^{-i\omega t}$
time harmonic variation. $\mathbf{e}_{x}$ is the unit vector in the
$x$ direction, and $E_{0}$ is the electric field amplitude, $k$
is the wave vector in free space. For a closed two level $J=0\rightarrow J=1$
atomic transition~\cite{Bettles2016,lambropoulos2007},
the atomic response is isotropic and linear (far from saturation).
The electric polarizability of each atom is defined as $\alpha=\frac{-\frac{\Gamma_{0}}{2}\alpha_{0}}{\delta+i\frac{\left(\Gamma_{0}+\Gamma_{nr}\right)}{2}}$
where $\delta=\omega-\omega_{a}\ll\omega_{a}$ is the detuning frequency
between the light beam and the transition frequency of the atom and
$\alpha_{0}=\frac{6\pi}{k^{3}}$. We assume elastic scattering events and therefore the non-radiative decay is zero, i.e. $\Gamma_{nr}=0$.
The induced displacement volume current density for the atomic dimer can be written as
\begin{eqnarray}
\mathbf{J}\left(\mathbf{r},\omega\right) & = & -i\omega\left[\hat{\mathbf{p}}_{d}\left(\mathbf{r}_{d}\right)\delta\left(\mathbf{r}-\mathbf{r}_{d}\right)+\hat{\mathbf{p}}_{u}\left(\mathbf{r}_{u}\right)\delta\left(\mathbf{r}-\mathbf{r}_{u}\right)\right],\label{eq:Current}
\end{eqnarray}
where $\delta\left(\mathbf{r}-\mathbf{r}_{i}\right)$ is the Dirac
delta function, $\epsilon_{0}$ is the permittivity of the free space.
$\hat{\mathbf{p}}_{i}\left(\mathbf{r}_{i}\right)$ $\left(i=u\,\mathrm{and}\,d\right)$
is the induced electric dipole moment of the upper and lower dipole moments,
respectively and for the two atoms under consideration reads {[}Fig.~\ref{fig:Geometry_dimer}
(b){]}~\cite{Abajo:2007}
\begin{eqnarray}
\left[\begin{array}{c}
\hat{p}_{d}\\
\hat{p}_{u}
\end{array}\right] & = & \left[\begin{array}{cc}
\frac{1}{\epsilon_{0}\alpha} & -G_{EE}^{xx}\left(\mathbf{r}_{d},\mathbf{r}_{u}\right)\\
-G_{EE}^{xx}\left(\mathbf{r}_{u},\mathbf{r}_{d}\right) & \frac{1}{\epsilon_{0}\alpha}
\end{array}\right]^{-1}\left[\begin{array}{c}
E_{0}e^{-ikl/2}\\
E_{0}e^{ikl/2}
\end{array}\right].\label{eq:ED_Matrix}
\end{eqnarray}
Note that the induced moments have only an $x$-component ($\hat{\mathbf{p}}_{u/d}=\hat{p}_{u/d}\mathbf{e}_{x}$)
due to an $x$-polarized plane wave illumination. $G_{EE}^{xx}\left(\mathbf{r}_{d},\mathbf{r}_{u}\right)=G_{EE}^{xx}\left(\mathbf{r}_{u},\mathbf{r}_{d}\right)$
is the Green function {[}see the Green function section{]}
\begin{equation}
G_{EE}^{xx}\left(\mathbf{r}_{d},\mathbf{r}_{u}\right)=\frac{3}{2\alpha_{0}\epsilon_{0}}e^{i\zeta}\left(\frac{1}{\zeta}-\frac{1}{\zeta^{3}}+\frac{i}{\zeta^{2}}\right),\,\,\,\zeta=k\left|\mathbf{r}_{d}-\mathbf{r}_{u}\right|=kl.
\end{equation}
Eq.~\ref{eq:ED_Matrix} can be simplified as
\begin{eqnarray}
\hat{p}_{d} & = & \epsilon_{0}\alpha_{d}E_{0}=\epsilon_{0}\alpha\frac{e^{-ikl/2}+\epsilon_{0}\alpha G_{EE}^{xx}e^{ikl/2}}{\left(1-\epsilon_{0}\alpha G_{EE}^{xx}\right)\left(1+\epsilon_{0}\alpha G_{EE}^{xx}\right)}E_{0},\nonumber \\
\hat{p}_{u} & = & \epsilon_{0}\alpha_{u}E_{0}=\epsilon_{0}\alpha\frac{e^{ikl/2}+\epsilon_{0}\alpha G_{EE}^{xx}e^{-ikl/2}}{\left(1-\epsilon_{0}\alpha G_{EE}^{xx}\right)\left(1+\epsilon_{0}\alpha G_{EE}^{xx}\right)}E_{0},
\end{eqnarray}
where $\alpha_{d}$ and $\alpha_{u}$ defined as
\begin{eqnarray}
\alpha_{d} & = & \alpha\left[\frac{\mathrm{cos}\left(kl/2\right)}{D_{-}}-i\frac{\mathrm{sin}\left(kl/2\right)}{D_{+}}\right],\nonumber \\
\alpha_{u} & = & \alpha\left[\frac{\mathrm{cos}\left(kl/2\right)}{D_{-}}+i\frac{\mathrm{sin}\left(kl/2\right)}{D_{+}}\right],
\end{eqnarray}
where $D_{\pm}=1\pm\alpha G_{EE}^{xx}\left(\mathbf{r}_{d},\mathbf{r}_{u}\right)$
and $\alpha_{u/d}$ is related to effective polarizability of the
dimer. Using above expressions, Eq.~\ref{eq:Current} can be written
\begin{eqnarray}
\mathbf{J}\left(\mathbf{r}\right) & = & -i\omega\epsilon_{0}E_{0}\left[\alpha_{d}\delta\left(\mathbf{r}-\mathbf{r}_{d}\right)+\alpha_{u}\delta\left(\mathbf{r}-\mathbf{r}_{u}\right)\right]\mathbf{e}_{x},\label{eq:J_ind}
\end{eqnarray}
which is the induced displacement volume current density for the atomic
dimer when illuminated by a plane wave ( $\mathbf{E}_{\mathrm{inc}}=E_{0}e^{ikz}\mathbf{e}_{x}$).
Now, we can use the multipole expansion {[}see Ref.~\cite{Alaee:2018},
Table II{]} and Eq.~\ref{eq:J_ind} to obtain the induced multipole
moments of the atomic dimer at the center $\mathbf{r}=0$.
\begin{figure}
\begin{centering}
\includegraphics[width=13cm]{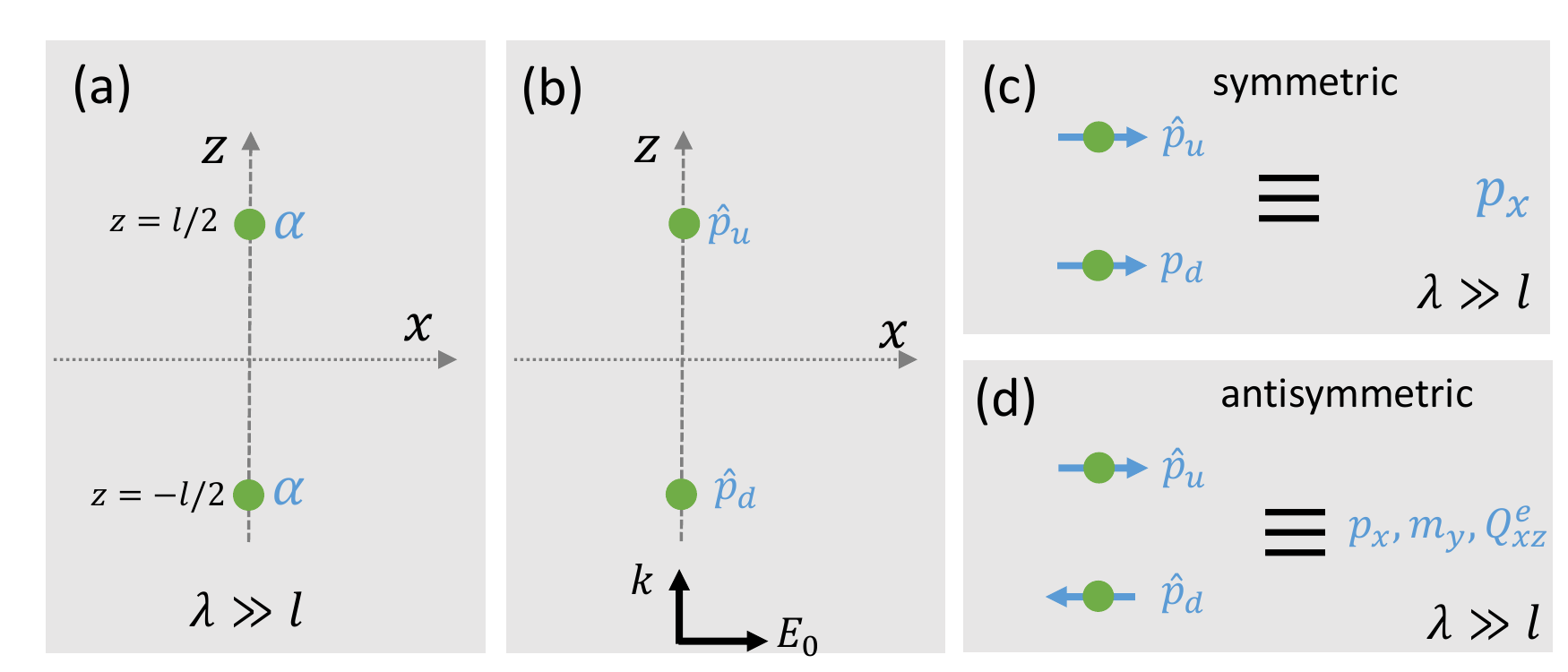}
\par\end{centering}
\caption{(a) An atomic dimer consisting of two identical atoms with electric
polarizability $\alpha$ placed at $\mathbf{r}_{u/d}=[0,0,\pm l/2]$.
(b) An atomic dimer when illuminated by an x-polarized plane wave
propagating in the $z$ direction, i.e. $\mathbf{E}_{\mathrm{inc}}=E_{0}e^{ikz}\mathbf{e}_{x}$
and the induced upper and lower electric dipole moments (i.e. $\hat{p}_{u}$,
$\hat{p}_{d}$), respectively. (c)-(d) Symmetric and antisymmetric
modes and their equivalent multipole moments obtained by applying
the multipole expansion at the center of the atomic dimer $\mathbf{r}=0$.
Note that the magnetic quadrupole moment can be neglected for $\lambda\gg l$.\label{fig:Geometry_dimer}}
\end{figure}
The induced \textit{effective} electric dipole moment of the atomic
dimer by applying the multipole expansion at the center of the dimer
$\mathbf{r}=0$ {[}Fig.~\ref{fig:Geometry_dimer} (b)-(d){]} read
as~\cite{Alaee:2018}
\begin{eqnarray}
p_{\beta} & = & -\frac{1}{i\omega}\left\{ \int dvJ_{\beta}j_{0}\left(kr\right)+\frac{k^{2}}{2}\int dv\left[3\left(\mathbf{r}\cdot\mathbf{J}\right)r_{\beta}-r^{2}J_{\beta}\right]\frac{j_{2}\left(kr\right)}{\left(kr\right)^{2}}\right\} ,\label{eq:p_ME}
\end{eqnarray}
where $\beta=x,y,z$ and $j_{n}\left(kr\right)$ are spherical Bessel
functions. Now by substituting Eq.~\ref{eq:J_ind} into Eq.~\ref{eq:p_ME},
we have

\begin{eqnarray}
p_{x} & = & -\frac{1}{i\omega}\left\{ \int J_{x}j_{0}\left(kr\right)dv+\frac{k^{2}}{2}\int dv\left[3\left(\mathbf{r}\cdot\mathbf{J}\right)x-r^{2}J_{x}\right]\frac{j_{2}\left(kr\right)}{\left(kr\right)^{2}}\right\} ,\nonumber \\
 & = & -\frac{1}{i\omega}\int J_{x}\left[j_{0}\left(kr\right)-\frac{1}{2}j_{2}\left(kr\right)\right]dv,\nonumber \\
 & = & \epsilon_{0}E_{0}\left(\alpha_{d}+\alpha_{u}\right)\left[j_{0}\left(kl/2\right)-\frac{j_{2}\left(kl/2\right)}{2}\right].\nonumber \\
 & = & \epsilon_{0}E_{0}\frac{2\alpha}{D_{-}}\left[j_{0}\left(kl/2\right)-\frac{j_{2}\left(kl/2\right)}{2}\right]\mathrm{cos}\left(kl/2\right).\label{eq:dipole_dimer}
\end{eqnarray}

where $\mathbf{r}\cdot\mathbf{J}=0$, $\mathbf{J}\left(\mathbf{r}\right)=J_{x}(\mathbf{r})\mathbf{e}_{x}$
and $\mathbf{r}=\mathbf{r}_{u/d}=\pm l/2\mathbf{e}_{z}$. Note that
$y$ and $z$ components of the electric dipole moments of the dimer
are zero, i.e. $p_{y}=0,$ and $p_{z}=0$ {[}see Eq.~\ref{eq:J_ind}
and the illumination direction in Fig.~\ref{fig:Geometry_dimer}
(b){]}. Next by using the definition of the electric dipole moment,
i.e. $p_{x}=\epsilon_{0}\alpha_{\mathrm{ed}}E_{0}$, the \textit{effective}
electric polarizability of the atomic dimer can be defined

\begin{equation}
\boxed{\alpha_{\mathrm{ed}}=\frac{2\alpha}{D_{-}}\left[j_{0}\left(kl/2\right)-\frac{j_{2}\left(kl/2\right)}{2}\right]\mathrm{cos}\left(kl/2\right)}
\end{equation}
The induced \textit{effective} magnetic dipole moment at the center
of the dimer $\mathbf{r}=0$ {[}Fig.~\ref{fig:Geometry_dimer} (b)-(d){]}
read as~\cite{Alaee:2018}
\begin{eqnarray}
m_{\beta} & = & \frac{3}{2}\int dv\left(\mathbf{r}\times\mathbf{J}\right)_{\beta}\frac{j_{1}\left(kr\right)}{kr},\label{eq:m_ME}
\end{eqnarray}
where $\beta=x,y,z$, by substituting Eq.~\ref{eq:J_ind} into Eq.~\ref{eq:m_ME},
we obtain
\begin{eqnarray}
m_{y} & = & \frac{3}{2}\int dv\left(\mathbf{r}\times\mathbf{J}\right)_{y}\frac{j_{1}\left(kr\right)}{kr}=\frac{3}{2}\int\left(zJ_{x}-xJ_{z}\right)\frac{j_{1}\left(kr\right)}{kr}dv,\nonumber \\
 & = & \frac{3}{2k}\int zJ_{x}\frac{j_{1}\left(kr\right)}{r}dv,\nonumber \\
 & = & \frac{3i}{2}\frac{E_{0}}{Z_{0}}\left(\alpha_{d}-\alpha_{u}\right)j_{1}\left(kl/2\right),\nonumber \\
 & = & \left[\frac{3\alpha}{D_{+}}j_{1}\left(kl/2\right)\mathrm{sin}\left(kl/2\right)\right]\frac{E_{0}}{Z_{0}}.\label{eq:my_dimer}
\end{eqnarray}
where $Z_{0}=\sqrt{\frac{\mu_{0}}{\epsilon_{0}}}$ is the intrinsic
impedance of the free space. Note that we have only the \textit{y}-component
of the magnetic moment. Now, by using the definition of the magnetic
dipole moment, i.e. $m_{y}=\alpha_{\mathrm{md}}H_{0}=\alpha_{\mathrm{md}}E_{0}/Z_{0}$,
the \textit{effective} magnetic polarizability can be defined
\begin{equation}
\boxed{\alpha_{\mathrm{md}}=\frac{3\alpha}{D_{+}}j_{1}\left(kl/2\right)\mathrm{sin}\left(kl/2\right)}
\end{equation}
The induced \textit{effective} electric quadrupole moment at the center
of the dimer $\mathbf{r}=0$ {[}Fig.~\ref{fig:Geometry_dimer} (b)-(d){]}
read as~\cite{Alaee:2018}

\begin{equation}
\begin{alignedat}{1}Q_{\mu\nu}^{\mathrm{e}} & =-\frac{3}{i\omega}\left\{ \int dv\left[3\left(r_{\nu}J_{\mu}+r_{\mu}J_{\nu}\right)-2\left(\mathbf{r}\cdot\mathbf{J}\right)\delta_{\mu\nu}\right]\frac{j_{1}\left(kr\right)}{kr}\right.\\
 & \left.+2k^{2}\int dv\left[5r_{\mu}r_{\nu}\left(\mathbf{r}\cdot\mathbf{J}\right)-\left(r_{\mu}J_{\nu}+r_{\nu}J_{\mu}\right)r^{2}-r^{2}\left(\mathbf{r}\cdot\mathbf{J}\right)\delta_{\mu\nu}\right]\frac{j_{3}\left(kr\right)}{\left(kr\right)^{3}}\right\} ,
\end{alignedat}
\label{eq:Qe_ME}
\end{equation}
where $\mu,\nu=x,y,z$, and $\delta_{\mu\nu}$ is the Kronecker delta.
Next, by substituting Eq.~\ref{eq:J_ind} into Eq.~\ref{eq:Qe_ME},
we have

\begin{equation}
\begin{alignedat}{1}Q_{xz}^{\mathrm{e}} & =-\frac{3}{i\omega}\left[\int dv3\left(zJ_{x}\right)\frac{j_{1}\left(kr\right)}{kr}-2\int dv\left(zJ_{x}\right)\frac{j_{3}\left(kr\right)}{kr}\right],\\
 & =-\frac{3}{i\omega}\int dv\frac{zJ_{x}}{kr}\left[3j_{1}\left(kl/2\right)-2j_{3}\left(kl/2\right)\right],\\
 & =\frac{3}{k}\epsilon_{0}E_{0}\left(\alpha_{u}-\alpha_{d}\right)\left[3j_{1}\left(kl/2\right)-2j_{3}\left(kl/2\right)\right],\\
 & =\frac{6i}{k}\frac{\alpha}{D_{+}}\epsilon_{0}E_{0}\left[3j_{1}\left(kl/2\right)-2j_{3}\left(kl/2\right)\right]\mathrm{sin}\left(kl/2\right).
\end{alignedat}
\label{eq:Qe_dimer}
\end{equation}

Note that other components of the tensor in Eq. \ref{eq:Qe_ME} are
zero. Thus, by using the definition of the electric quadrupole moment~\cite{alu2009},
$Q_{xz}^{\mathrm{e}}=\frac{1}{2}\epsilon_{0}\alpha_{\mathrm{eq}}\left(\frac{\partial E_{x}}{\partial z}+\frac{\partial E_{z}}{\partial x}\right)=\frac{ik}{2}\epsilon_{0}E_{0}\alpha_{\mathrm{eq}}$,
the \textit{effective} electric quadrupole polarizability can be defined

\begin{equation}
\boxed{\alpha_{\mathrm{eq}}=\frac{12}{k^{2}}\frac{\alpha}{D_{+}}\left[3j_{1}\left(kl/2\right)-2j_{3}\left(kl/2\right)\right]\mathrm{sin}\left(kl/2\right)}
\end{equation}

The induced \textit{effective} magnetic quadrupole moment at the center
of the dimer $\mathbf{r}=0$ {[}Fig.~\ref{fig:Geometry_dimer} (b)-(d){]}
read as~\cite{Alaee:2018}

\begin{eqnarray}
Q_{\mu\nu}^{m} & = & 15\int dv\left\{ r_{\mu}\left(\mathbf{r}\times\mathbf{J}\right)_{\nu}+r_{\nu}\left(\mathbf{r}\times\mathbf{J}\right)_{\mu}\right\} \frac{j_{2}\left(kr\right)}{\left(kr\right)^{2}},\label{eq:Qm_ME}
\end{eqnarray}
where $\mu,\nu=x,y,z$, by substituting Eq.~\ref{eq:J_ind} on Eq.~\ref{eq:Qm_ME},
we obtain
\begin{eqnarray}
Q_{zy}^{m} & = & 15\int dv\left[z^{2}J_{x}\right]\frac{j_{2}\left(kr\right)}{\left(kr\right)^{2}},\nonumber \\
 & = & \frac{15}{ik}\frac{E_{0}}{Z_{0}}\left(\alpha_{u}+\alpha_{d}\right)j_{2}\left(kl/2\right),\nonumber \\
 & = & \frac{30}{ikD_{-}}\frac{E_{0}}{Z_{0}}\alpha j_{2}\left(kl/2\right)\mathrm{cos}\left(kl/2\right).
\end{eqnarray}

Thus, by using the definition of the magnetic quadrupole moment~\cite{alu2009},
i.e. $Q_{zy}^{\mathrm{m}}=\frac{1}{2}\alpha_{\mathrm{mq}}\left(\frac{\partial H_{z}}{\partial y}+\frac{\partial H_{y}}{\partial z}\right)=\frac{ik}{2}\frac{E_{0}}{Z_{0}}\alpha_{\mathrm{mq}}$,
the \textit{effective} electric quadrupole polarizability read as

\begin{equation}
\boxed{\alpha_{\mathrm{mq}}=-\frac{60}{k^{2}}\frac{\alpha}{D_{-}}j_{2}\left(kl/2\right)\mathrm{cos}\left(kl/2\right)}
\end{equation}

\subsection{Scattering cross section}

In this section, we derive an expression to calculate the scattering
cross section of an atomic dimer. Using the induced multipole moments,
we can obtain the scattering cross section of the atomic dimer~\cite{Alaee:2018} 
\begin{eqnarray}
C_{\mathrm{sca}} & \approx & \frac{k^{4}}{6\pi\epsilon_{0}^{2}\left|E_{0}\right|^{2}}\left(\left|p_{x}\right|^{2}+\left|\frac{m_{y}}{c}\right|^{2}+\frac{3}{5}\left|\frac{k}{6}Q_{xz}^{e}\right|^{2}+\frac{3}{5}\left|\frac{k}{6c}Q_{zy}^{m}\right|^{2}\right).\label{eq:CSca_ME1}
\end{eqnarray}

Eq.~\ref{eq:CSca_ME1} can be written as a function of polarizabilities

\begin{eqnarray}
C_{\mathrm{sca}} & = & \frac{k^{4}}{6\pi}\left[\left|\alpha_{\mathrm{ed}}\right|^{2}+\left|\alpha_{\mathrm{md}}\right|^{2}+\frac{3}{5}\left|\frac{k^{2}}{12}\alpha_{\mathrm{eq}}\right|^{2}+\frac{3}{5}\left|\frac{k^{2}}{12}\alpha_{\mathrm{mq}}\right|^{2}\right],\label{eq:Csca_ME}
\end{eqnarray}
where we used the electric and magnetic multipole moments definitions
$p_{x}=\epsilon_{0}\alpha_{\mathrm{ed}}E_{0}$, $m=\alpha_{\mathrm{md}}H_{0}$,
$Q_{xz}^{e}=Q_{zx}^{e}=\frac{ik}{2}\alpha_{\mathrm{eq}}E_{0}$ and $Q_{zy}^{\mathrm{m}}=Q_{yz}^{\mathrm{m}}=\frac{ik}{2}\frac{E_{0}}{Z_{0}}\alpha_{\mathrm{mq}}$.
For the atomic dimer with $\lambda\gg l$, magnetic quadrupole moment
is negligible, i.e. $Q_{zy}^{m}\approx0$. Note that Eq.~\ref{eq:Csca_ME}
is not sufficient for $\lambda\ll l$ and one should consider higher
order multipole moments.
An alternative approach to obtain the scattering cross section of
the atomic dimer is based on the coupled dipole theory [see the
coupled dipole theory section]. We assume that the nonradiative
losses is zero in the atomic dimer. Thus, according to the optical theorem, the extinction cross section is identical to the scattering cross section, i.e. $C_{\mathrm{sca}}=C_{\mathrm{ext}}$, and therefore,
\begin{eqnarray}
C_{\mathrm{ext}} & = & \frac{k}{\epsilon_{0}\left|E_{0}\right|^{2}}\mathrm{Im}\left[p_{d}E_{\mathrm{inc}}^{*}\left(\mathbf{r}_{d}\right)+p_{u}E_{\mathrm{inc}}^{*}\left(\mathbf{r}_{u}\right)\right],\nonumber \\
 & = & k\mathrm{Im}\left[\alpha_{d}e^{ikl/2}+\alpha_{u}e^{-ikl/2}\right]\\
 & = & k\mathrm{Im}\left[2\alpha\frac{1+\epsilon_{0}\alpha\mathrm{G_{EE}^{xx}cos}\left(kl/2\right)}{D_{-}D_{+}}\right].\label{eq:Cext_exact}
\end{eqnarray}
Eq.~\ref{eq:Csca_ME} is identical to the Eq.~$\ref{eq:Cext_exact}$
if an atomic dimer is without nonradiative losses and $\lambda\gg l$.

\subsection{Radiation pattern}

Radiation pattern of an atomic dimer can be find by using the radiated
far field~\cite{Jackson1999,campione2015,Alaee_kerker:15}
\begin{eqnarray}
\mathbf{E}_{ED} & = & \frac{k^{2}}{4\pi\epsilon_{0}}\frac{e^{ikr}}{r}p_{x}\left(-\mathrm{sin}\varphi\mathbf{e}_{\varphi}+\mathrm{cos}\theta\mathrm{cos}\varphi\mathbf{e}_{\theta}\right),\nonumber \\
\mathbf{E}_{MD} & = & \frac{k^{2}}{4\pi\epsilon_{0}}\frac{e^{ikr}}{r}\frac{m_{y}}{c}\left(-\mathrm{cos}\theta\mathrm{sin}\varphi\mathbf{e}_{\varphi}+\mathrm{cos}\varphi\mathbf{e}_{\theta}\right),\nonumber \\
\mathbf{E}_{EQ} & = & \frac{k^{2}}{4\pi\epsilon_{0}}\frac{e^{ikr}}{r}\frac{ik}{6}Q_{zx}^{e}\left[\mathrm{cos}\theta\mathrm{sin}\varphi\mathbf{e}_{\varphi}-\mathrm{\left(\mathrm{2cos^{2}}\theta-1\right)cos}\varphi\mathbf{e}_{\theta}\right],\nonumber \\
\mathbf{E}_{MQ} & = & \frac{k^{2}}{4\pi\epsilon_{0}}\frac{e^{ikr}}{r}\frac{ik}{6c}Q_{zy}^{m}\left[\left(\mathrm{2cos^{2}}\theta-1\right)\mathrm{sin}\varphi\mathbf{e}_{\varphi}-\mathrm{\mathrm{cos}\theta cos}\varphi\mathbf{e}_{\theta}\right],
\end{eqnarray}

where $r,\theta,\varphi$ are the radial distance, polar angle, and
azimuthal angle. In the xz-plane, i.e. $\varphi=0$, the radiation
pattern considering the contribution from all multipole moments (up
to magnetic quadrupole) can be written as

\begin{eqnarray}
\mathbf{E} & \approx & \frac{k^{2}}{4\pi\epsilon_{0}}\frac{e^{ikr}}{r}\left[p_{x}\mathrm{cos}\theta+\frac{m_{y}}{c}-\frac{ik}{6}Q_{xz}^{e}\left(\mathrm{2cos^{2}}\theta-1\right)-\frac{ik}{6c}Q_{zy}^{m}\mathrm{cos}\theta\right]\mathbf{e}_{\theta},
\end{eqnarray}
Let us consider an atomic dimer with $\lambda\gg l$ that supports
symmetric and antisymmetric modes. At the symmetric mode resonance
frequency~{[}see Fig.~\ref{fig:Geometry_dimer} (c){]}, $\left|p_{x}\right|\gg\left|\frac{m_{y}}{c}\right|$,
and $\left|p_{x}\right|\gg\left|\frac{ik}{6}Q_{xz}^{e}\right|$ and
the magnetic quadrupole moment is negligible, i.e. $Q_{zy}^{m}\approx0$.
Thus, the radiation pattern is similar to a pure electric dipole moment,
i.e.
\begin{eqnarray}
\mathbf{E} & \approx & \frac{k^{2}}{4\pi\epsilon_{0}}\frac{e^{ikr}}{r}p_{x}\mathrm{cos}\theta\mathbf{e}_{\theta},
\end{eqnarray}
where $p_{x}$ for the atomic dimer can be obtained using Eq.~\ref{eq:dipole_dimer}.
In the following section, we will show that an atomic bilayer composed
of atomic dimer act as an atomic electric mirror at the symmetric
mode resonance frequency. However, at the antisymmetric mode resonance
frequency, $\left|p_{x}\right|\ll\left|\frac{m_{y}}{c}\right|$, and
$\left|p_{x}\right|\ll\left|\frac{ik}{6}Q_{xz}^{e}\right|$ the radiation
pattern is similar to a superposition of magnetic dipole and electric
quadrupole moments, i.e.
\begin{eqnarray}
\mathbf{E} & \approx & \frac{k^{2}}{4\pi\epsilon_{0}}\frac{e^{ikr}}{r}\left[\frac{m_{y}}{c}-\frac{ik}{6}Q_{xz}^{e}\left(\mathrm{2cos^{2}}\theta-1\right)\right]\mathbf{e}_{\theta},\nonumber \\
 & \approx & \frac{k^{2}}{4\pi\epsilon_{0}}\frac{e^{ikr}}{r}\frac{m_{y}}{c}\left(\mathrm{2cos^{2}}\theta\right)\mathbf{e}_{\theta},\label{eq:E_AsyMode}
\end{eqnarray}
where $m_{y}$ and $Q_{xz}^{e}$ for the atomic dimer can be obtained
using Eq.~\ref{eq:my_dimer} and Eq.~\ref{eq:Qe_dimer}, respectively~{[}see
Fig.~\ref{fig:Geometry_dimer} (d){]}. Note that for $\lambda\gg l$,
we can show that $\frac{m_{y}}{c}\approx-\frac{ik}{6}Q_{xz}^{e}$
which is used in the derivation of Eq.~\ref{eq:E_AsyMode}. The radiation
patterns of both modes are plotted in Fig.~\ref{fig:RadiationPattern}.
It can be seen that the symmetric mode resonance has an in phase radiation
pattern in both forward and backward directions. Whereas for the anti-symmetric
mode the radiated fields are out of phase in forward and backward
directions. In the following section, we show that an atomic
bilayer consist of atomic dimer acts as an electric or a magnetic
mirror for symmetric or antisymmetric modes, respectively~{[}see
Fig.~\ref{fig:Geometry_dimer} and Fig.~\ref{fig:GaussianResults}{]}.
\begin{figure}
\begin{centering}
\includegraphics[width=13cm]{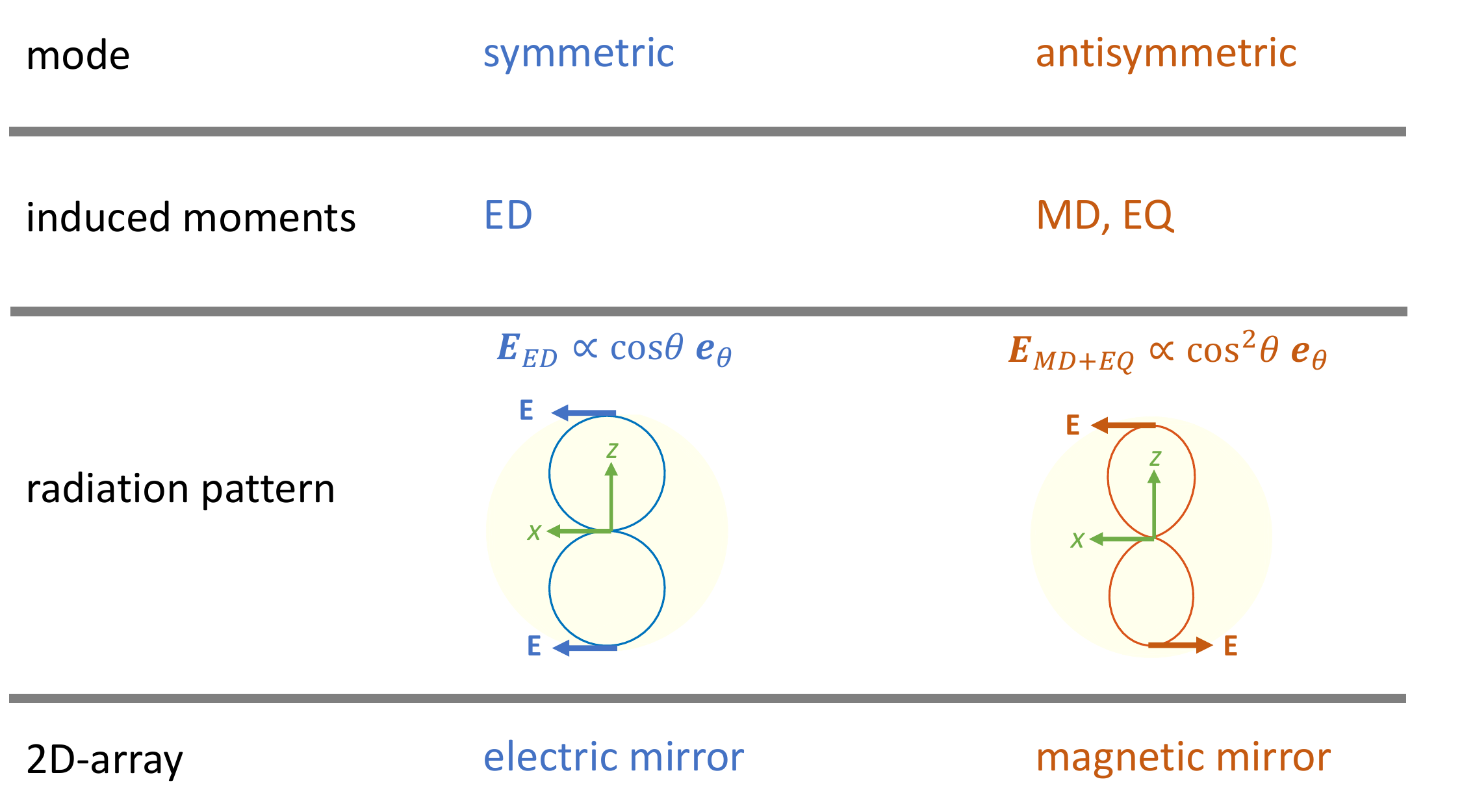}
\par\end{centering}
\caption{An atomic dimer or a bilayer support both symmetric and antisymmetric
modes. The symmetric mode acts as an electric mirror whereas the antisymmetric
mode acts as a magnetic mirror.\label{fig:RadiationPattern}}
\end{figure}
\section{Atomic monolayer metasurface (AMM): atomic electric mirrors}
\subsection{Reflection and transition coefficients}
Let us consider an atomic monolayer composed of atoms with only electric
dipole transition moments. The atoms are periodically arranged in
$xy$-plane at $z=0$, $\mathbf{r}_{n}=\mathbf{r}_{n_{x},n_{y}}=\left(n_{x}\mathbf{e}_{x}+n_{y}\mathbf{e}_{y}\right)\Lambda$.
$n_{x}$ and $n_{y}$ are integer numbers and $\Lambda$ is the periodicity
in $x$ and $y$ directions. The reflected and transmitted electric
fields ($\mathbf{E}_{r}$ and $\mathbf{E}_{t}$) at the interface
when illuminated by a polarized plane wave propagating in the $z$
direction, i.e. $\mathbf{E}_{\mathrm{inc}}=E_{0}e^{ikz}\mathbf{e}_{x}$
defined as~\cite{Tretyakov:03,Abajo:2007,shahmoon2017,Alaee:2017Review}
\begin{eqnarray}
\mathbf{E}_{r}=-\frac{1}{2}Z_{0}\mathbf{J}_{e},\,\,\,\,\,\,\,\,\,\mathbf{E}_{t} & = & \mathbf{E}_{\rm inc}-\frac{1}{2}Z_{0}\mathbf{J}_{e},\label{eq:Er_Et_ML}
\end{eqnarray}
where $Z_{0}=\sqrt{\frac{\mu_{0}}{\epsilon_{0}}}$ is the impedance
of the free space. $\mathbf{J}_{e}$ is the induced averaged surface
electric current. $\mathbf{p}$ is the effective induced electric
dipole moment of the atomic array and defined as
\begin{eqnarray}
\mathbf{p}\left(\mathbf{r}_{0}\right) & = & \epsilon_{0}\alpha\left[\mathbf{E}_{\mathrm{inc}}+\underset{n,\,n\neq0}{\sum}\mathbf{\bar{\bar{G}}}_{EE}\left(\mathbf{r}_{0},\mathbf{r}_{n}\right)\cdot\mathbf{p}\left(\mathbf{r}_{0}\right)\right],
\end{eqnarray}
and using $\mathbf{E}_{\mathrm{inc}}=E_{0}e^{ikz}\mathbf{e}_{x}$, we get
\begin{eqnarray}
\mathbf{p}\left(\mathbf{r}_{0}\right) & = & \epsilon_{0}E_{0}\alpha_{\mathrm{eff}}\mathbf{e}_{x},\,\,\,\,\alpha_{\mathrm{eff}}=\frac{\alpha}{1-\epsilon_{0}\alpha\underset{n,\,n\neq0}{\sum}G_{EE}^{xx}\left(\mathbf{r}_{0},\mathbf{r}_{n}\right)},
\end{eqnarray}
where $\mathbf{r}_{0}=0$, $\underset{n\neq0}{\sum}G_{EE}^{xx}\left(\mathbf{r}_{0},\mathbf{r}_{n}\right)$
is the interaction constant (or the l\textit{attice sum} of the dipolar
interaction tensor). Now by substituting the induced averaged surface
electric current $\mathbf{J}_{e}=-i\omega\frac{\mathbf{p}}{\Lambda^{2}}$
into Eq.~\ref{eq:Er_Et_ML}, the reflection coefficient $r=\frac{E_{r}}{E_{0}}$
for an atomic monolayer metasurface can be obtained~\cite{Tretyakov:03,Abajo:2007,Alaee:2017Review}

\begin{eqnarray}
r & = & \frac{ik}{2\Lambda^{2}}\left(\frac{1}{\frac{1}{\alpha}-\epsilon_{0}\underset{n,\,n\neq0}{\sum}G_{EE}^{xx}\left(0,\mathbf{r}_{n}\right)}\right)=\frac{ik}{2\Lambda^{2}}\alpha_{\mathrm{eff}},\nonumber \\
 & = & \frac{ik}{2\Lambda^{2}}\left(\frac{-\frac{\Gamma_{0}}{2}\alpha_{0}}{\delta+i\frac{\Gamma_{0}}{2}+\frac{\Gamma_{0}}{2}\alpha_{0}\epsilon_{0}\underset{n,\,n\neq0}{\sum}G_{EE}^{xx}\left(0,\mathbf{r}_{n}\right)}\right)\nonumber \\
 & = & \frac{ik}{2\Lambda^{2}}\left(\frac{-\alpha_{0}\frac{\Gamma_{0}}{2}}{\delta-\Delta+i\left(\frac{\Gamma_{0}}{2}+\frac{\Gamma}{2}\right)}\right),\label{eq:r_ML}
\end{eqnarray}
where $\Delta$ and $\Gamma$ defined as
\begin{equation}
\Delta=-\epsilon_{0}\alpha_{0}\frac{\Gamma_{0}}{2}\mathrm{Re}\left[\underset{n,\,n\neq0}{\sum}G_{EE}^{xx}\left(0,\mathbf{r}_{n}\right)\right],\,\,\,\,\,\,\,\,\,\frac{\Gamma}{2}=\epsilon_{0}\alpha_{0}\frac{\Gamma_{0}}{2}\mathrm{Im}\left[\underset{n,\,n\neq0}{\sum}G_{EE}^{xx}\left(0,\mathbf{r}_{n}\right)\right],
\end{equation}
and transmission coefficient can be calculated by $t=1+r$. The reflection
coefficient can be written as
\begin{eqnarray}
r & = & \frac{-\frac{ik}{2\Lambda^{2}}\alpha_{0}\frac{\Gamma_{0}}{2}}{\delta-\Delta+i\frac{k}{2\Lambda^{2}}\alpha_{0}\frac{\Gamma_{0}}{2}},\label{eq:r_monolayer}
\end{eqnarray}
where $\mathrm{Im}\left[\underset{n,\,n\neq0}{\sum}G_{EE}^{xx}\left(0,\mathbf{r}_{n}\right)\right]=-\frac{1}{\epsilon_{0}\alpha_{0}}+\frac{k}{2\Lambda^{2}\epsilon_{0}}$
which is obtained in the next section using the energy conservation.
Note that the real part of the interaction constant (or lattice sum),
i.e. $\mathrm{Re}\left[\underset{n,\,n\neq0}{\sum}G_{EE}^{xx}\left(0,\mathbf{r}_{n}\right)\right]$
can be calculated numerically~\cite{Tretyakov:03,Abajo:2007,Alaee:2017Review}.

\subsection{Energy conservation and interaction constant}

The conservation of energy can be used to obtain an exact expression
for the imaginary part of the the interaction constant (lattice sum),
i.e. $\mathrm{Im}\left[C\right]=\mathrm{Im}\left[\underset{n\neq0}{\sum}G_{EE}^{xx}\left(\mathbf{r}=\mathbf{0},\mathbf{r}_{n}\right)\right]$,
we assume the nonradiative losses is zero in the atomic monolayer
and have
\begin{eqnarray}
A=1-R-T & = & 1-\left|r\right|^{2}-\left|1+r\right|^{2}=0\label{eq:TR_EC}
\end{eqnarray}
where $T=\left|1+r\right|^{2}$ , and $R=\left|r\right|^{2}$ are
the transmission and reflection from the atomic monolayer. Using Eq.~\ref{eq:TR_EC},
it can be shown that $\frac{\mathrm{Re}\left(r\right)}{\left|r\right|^{2}}=\mathrm{Re}\left[\frac{1}{r}\right]=-1$
and we get
\begin{eqnarray}
\mathrm{Re}\left[\frac{1}{r}\right] & = & -1,\nonumber \\
\mathrm{Re}\left(\frac{1}{\frac{ik}{2\Lambda^{2}}\alpha_{\mathrm{eff}}}\right) & = & -1,\nonumber \\
\mathrm{Im}\left(\frac{1}{\alpha_{\mathrm{eff}}}\right) & = & -\frac{k}{2\Lambda^{2}}.\label{eq:EC_E1}
\end{eqnarray}
Now by using the definition of the effective polarizability $\alpha_{\mathrm{eff}}=\frac{\alpha}{1-\alpha\epsilon_{0}\underset{n,\,n\neq0}{\sum}G_{EE}^{xx}\left(0,\mathbf{r}_{n}\right)}$,
we have
\begin{eqnarray}
\mathrm{Im}\left[\frac{1}{\alpha_{\mathrm{eff}}}\right] & = & \mathrm{Im}\left(\frac{1}{\alpha}\right)-\mathrm{Im}\left[\epsilon_{0}\underset{n,\,n\neq0}{\sum}G_{EE}^{xx}\left(0,\mathbf{r}_{n}\right)\right].\label{eq:EC_E2}
\end{eqnarray}
From the definition of polarizability $\alpha=\frac{-\frac{\Gamma_{0}}{2}\alpha_{0}}{\delta+i\frac{\Gamma_{0}}{2}}$,
we obtain $\mathrm{Im}\left(\frac{1}{\alpha}\right)=-\frac{1}{\alpha_{0}}=\frac{k^{3}}{6\pi}$.
Finally, by using Eqs.~(\ref{eq:EC_E1}) and (\ref{eq:EC_E2}), the
exact expression for the imaginary part of the interaction constant,
i.e. $\mathrm{Im}\left[\epsilon_{0}\underset{n,\,n\neq0}{\sum}G_{EE}^{xx}\left(0,\mathbf{r}_{n}\right)\right]$
can be obtained as~\cite{Tretyakov:03,Abajo:2007,Alaee:2017Review}
\begin{equation}
\mathrm{Im}\left[\underset{n,\,n\neq0}{\sum}G_{EE}^{xx}\left(0,\mathbf{r}_{n}\right)\right]=-\frac{1}{\epsilon_{0}\alpha_{0}}+\frac{k}{2\Lambda^{2}\epsilon_{0}}.
\end{equation}
\section{Atomic bilayer metasurface (ABM): atomic electric and magnetic mirrors}
Let us consider an atomic bilayer composed of atoms with \textit{only}
electric dipole transition moments~(see Fig.~\ref{fig:Geometry_dimer})~\cite{Tretyakov:03}.
Atoms are periodically arranged in two layers in $xy$-planes, the
position of the upper layer $\mathbf{r}_{u,n}=n_{x}\Lambda\mathbf{e}_{x}+n_{y}\Lambda\mathbf{e}_{y}+\frac{l}{2}\mathbf{e}_{z}$
and the position of the lower $\mathbf{r}_{d,n}=n_{x}\Lambda\mathbf{e}_{x}+n_{y}\Lambda\mathbf{e}_{y}-\frac{l}{2}\mathbf{e}_{z}$.
$n_{x}$ and $n_{y}$ are integer numbers and $\Lambda$ is the periodicity
in $x$ and $y$ directions. The atomic bilayer is illuminated by
an $x$-polarized incident plane wave propagating in $z$-direction
$\mathbf{E}_{\mathrm{inc}}=E_{0}e^{ikz}\mathbf{e}_{x}$ . Using $\mathbf{E}_{r}=-\frac{1}{2}Z_{0}\mathbf{J}=\frac{ik}{2\Lambda^{2}\epsilon_{0}}\mathbf{p}$,
the incident and transmitted electric fields by the atomic bilayer
metasurface at $z=0$ read as
\begin{eqnarray}
E_{r} & = & \frac{ik}{2\Lambda^{2}\epsilon_{0}}\left[p_{d}^{\mathrm{eff}}e^{-ikl/2}+p_{u}^{\mathrm{eff}}e^{ikl/2}\right],\nonumber \\
E_{t} & = & E_{0}+\frac{ik}{2\Lambda^{2}\epsilon_{0}}\left[p_{d}^{\mathrm{eff}}e^{ikl/2}+p_{u}^{\mathrm{eff}}e^{-ikl/2}\right].\label{eq:Er_Et_BL}
\end{eqnarray}
where $p_{u}^{\mathrm{eff}}$ and $p_{d}^{\mathrm{eff}}$ are the
\textit{effective} (collective) upper and lower electric dipole moments
and $e^{\pm ikl/2}$ are the propagation terms. The reflection and
transmission coefficients are given by
\begin{eqnarray}
r & = & \frac{ik}{2\Lambda^{2}\epsilon_{0}E_{0}}\left(p_{d}^{\mathrm{eff}}e^{-ikl/2}+p_{u}^{\mathrm{eff}}e^{ikl/2}\right),\nonumber \\
t & = & 1+\frac{ik}{2\Lambda^{2}\epsilon_{0}E_{0}}\left(p_{d}^{\mathrm{eff}}e^{ikl/2}+p_{u}^{\mathrm{eff}}e^{-ikl/2}\right),\label{eq:TR_BL-SM}
\end{eqnarray}
and defined as
\begin{eqnarray}
p_{d}^{\mathrm{eff}} & = & \epsilon_{0}\alpha\left[E_{\mathrm{inc}}\left(\mathbf{r}_{d,0}\right)+\underset{n,\,n\neq0}{\sum}G_{EE}^{xx}\left(\mathbf{r}_{d,0},\mathbf{r}_{d,n}\right)p_{d}^{\mathrm{eff}}+\underset{n}{\sum}G_{EE}^{xx}\left(\mathbf{r}_{d,0},\mathbf{r}_{u,n}\right)p_{u}^{\mathrm{eff}}\right],\nonumber \\
p_{u}^{\mathrm{eff}} & = & \epsilon_{0}\alpha\left[E_{\mathrm{inc}}\left(\mathbf{r}_{u,0}\right)+\underset{n,\,n\neq0}{\sum}G_{EE}^{xx}\left(\mathbf{r}_{u,0},\mathbf{r}_{u,n}\right)p_{u}^{\mathrm{eff}}+\underset{n}{\sum}G_{EE}^{xx}\left(\mathbf{r}_{u,0},\mathbf{r}_{d,n}\right)p_{d}^{\mathrm{eff}}\right],\label{eq:TL_H1}
\end{eqnarray}
and can be written as
\begin{eqnarray}
\left[\begin{array}{c}
p_{d}^{\mathrm{eff}}\\
p_{u}^{\mathrm{eff}}
\end{array}\right] & = & \left[\begin{array}{cc}
\frac{1}{\epsilon_{0}\alpha}-C_{dd} & -C_{du}\\
-C_{ud} & \frac{1}{\epsilon_{0}\alpha}-C_{uu}
\end{array}\right]^{-1}\left[\begin{array}{c}
E_{\mathrm{inc}}\left(\mathbf{r}_{d,0}\right)\\
E_{\mathrm{inc}}\left(\mathbf{r}_{u,0}\right)
\end{array}\right].\label{eq:Induced_ED_eff-SM}
\end{eqnarray}
For identical upper and lower atoms, i.e. $\alpha$, the interaction
constants are symmetric, i.e. $C_{ud}=C_{du}$ and $C_{uu}=C_{dd}$
and defined as
\begin{eqnarray}
C_{dd} & = & \underset{n,\,n\neq0}{\sum}G_{EE}^{xx}\left(\mathbf{r}_{d,0},\mathbf{r}_{d,n}\right),\nonumber \\
C_{du} & = & \underset{n}{\sum}G_{EE}^{xx}\left(\mathbf{r}_{d,0},\mathbf{r}_{u,n}\right).
\end{eqnarray}
In general, the interaction constant can be numerically calculated.
For a lossless system, the imaginary part of the interaction constant
(lattice sum) can be exactly calculated by using the conservation
of energy~\cite{Tretyakov:03}
\begin{eqnarray}
\mathrm{Im}\left[C_{dd}\right] & = & -\frac{1}{\epsilon_{0}\alpha_{0}}+\frac{k}{2\Lambda^{2}\epsilon_{0}},\nonumber \\
\mathrm{Im}\left[C_{du}\right] & = & \frac{k}{2\Lambda^{2}\epsilon_{0}}\mathrm{cos}\left(kl\right).
\end{eqnarray}
The effective electric and magnetic polarizabilities of the atomic
bilayer can be defined as
\begin{eqnarray}
\alpha_{\mathrm{ed}}^{\mathrm{eff}} & = & \frac{p_{d}^{\mathrm{eff}}+p_{u}^{\mathrm{eff}}}{\epsilon_{0}E_{0}}\mathrm{cos}\left(kl/2\right),\nonumber \\
\alpha_{\mathrm{md}}^{\mathrm{eff}} & = & i\frac{p_{d}^{\mathrm{eff}}-p_{u}^{\mathrm{eff}}}{\epsilon_{0}E_{0}}\mathrm{sin}\left(kl/2\right).\label{eq:EM_polarizabilities-SM}
\end{eqnarray}

At the symmetric mode resonance frequency the induced dipoles for
both lower and upper layers are almost identical, i.e. $p_{d}^{\mathrm{eff}}\approx p_{u}^{\mathrm{eff}}${[}see
Fig. 3 of the main manuscript{]}, thus we get

\begin{eqnarray}
r & = & \frac{ik}{2\Lambda^{2}\epsilon_{0}E_{0}}p_{d}^{\mathrm{eff}}\left(e^{-ikl/2}+e^{ikl/2}\right),\nonumber \\
 & = & \frac{ik}{\Lambda^{2}\epsilon_{0}E_{0}}p_{d}^{\mathrm{eff}}\mathrm{cos}\left(kl/2\right),\,\,\,\mathrm{cos}\left(kl/2\right)\approx1,\,\,\,\Lambda=\frac{\lambda}{2}=\frac{\pi}{k},\nonumber \\
 & \approx & \left(\frac{p_{d}^{\mathrm{eff}}}{\alpha_{0}\epsilon_{0}E_{0}}\right)\frac{6i}{\pi},
\end{eqnarray}
the total reflection, i.e. $r\approx-1$ occurs when $p_{d}^{\mathrm{eff}}\approx p_{u}^{\mathrm{eff}}=\left(\alpha_{0}\epsilon_{0}E_{0}\right)i\frac{\pi}{6}$
{[}see Fig. 3 of the main manuscript{]}. However, at the antisymmetric
mode resonance frequency, i.e. $p_{d}^{\mathrm{eff}}\approx-p_{u}^{\mathrm{eff}}$~(see
Fig. 3 of the main manuscript), we have
\begin{eqnarray}
r & = & \frac{ik}{2\Lambda^{2}\epsilon_{0}E_{0}}p_{d}^{\mathrm{eff}}\left(e^{-ikl/2}-e^{ikl/2}\right),\nonumber \\
 & = & \frac{k}{\Lambda^{2}\epsilon_{0}E_{0}}p_{d}^{\mathrm{eff}}\mathrm{sin}\left(kl/2\right),\,\,\,\mathrm{sin}\left(kl/2\right)\approx kl/2,\,\,\,\Lambda=\frac{\lambda}{2}=\frac{\pi}{k},\nonumber \\
 & \approx & \left(\frac{p_{d}^{\mathrm{eff}}}{\alpha_{0}\epsilon_{0}E_{0}}\right)\frac{6l}{\lambda},
\end{eqnarray}
the total reflection, i.e. $r\approx1$ occurs when $p_{d}^{\mathrm{eff}}\approx-p_{u}^{\mathrm{eff}}=\left(\alpha_{0}\epsilon_{0}E_{0}\right)\frac{\lambda}{6l}$.

\section{Atomic monolayer/bilayer in a planar cavity}

\subsection{Planar cavity}

In this section, we consider a planar cavity consists of two identical
distributed Bragg reflector (DBR) mirrors {[}see Fig.~4 of the main
manuscript{]}. The mirrors are separated by a distance of $L_{c}$.
The optical properties of the system is described by the transfer
matrix product~\cite{Saleh1991}
\begin{eqnarray}
M_{\mathrm{Cavity}} & = & M_{\mathrm{DBR1}}M_{\mathrm{FS}}M_{\mathrm{DBR2}},\\
M_{\mathrm{FS}} & = & \left[\begin{array}{cc}
e^{i\varphi} & 0\\
0 & e^{-i\varphi}
\end{array}\right],\\
M_{\mathrm{DBR1}} & = & \frac{1}{t_{M}}\left[\begin{array}{cc}
t_{M}^{2}-r_{M,R}r_{M,L} & r_{M,R}\\
-r_{M,L} & 1
\end{array}\right],\\
M_{\mathrm{DBR2}} & = & \frac{1}{t_{M}}\left[\begin{array}{cc}
t_{M}^{2}-r_{M,R}r_{M,L} & r_{M,L}\\
-r_{M,R} & 1
\end{array}\right],
\end{eqnarray}
where $\varphi=knd$, $n$ is the refractive index of the spacer and
here is assumed to be the free space ($n=1$). $r_{M,L}$ and $r_{M,R}$
are reflection coefficients of the DBR mirrors when illuminated from
left and right, respectively. For the reciprocal DBR mirrors, the
transmission coefficients are identical, i.e. $t_{M}=t_{M,L}=t_{M,R}$.
The transmission and reflection coefficients of the planar cavity
are defined as
\begin{eqnarray}
t_{c} & = & \frac{t_{M}^{2}e^{ikL_{\mathrm{c}}}}{1-e^{2ikL_{\mathrm{c}}}r_{M,R}^{2}},\nonumber \\
r_{c,L} & = & r_{c,R}=-\frac{r_{M,L}+e^{2ikL_{\mathrm{c}}}r_{M,R}\left(t_{M}^{2}-r_{M,R}r_{M,L}\right)}{1-e^{2ikL_{\mathrm{c}}}r_{M,R}^{2}},
\end{eqnarray}
we assumed that there is no nonradiative losses, thus the cavity completely
transmits the light at $\omega=\omega_{c}$. The finesse of the planer
cavity can be defined as
\begin{equation}
F=\frac{\pi\left|\sqrt{r_{M,R}^{2}}\right|}{1-\left|r_{M,R}^{2}\right|}.
\end{equation}
\subsection{Atomic monolayer/bilayer metasurface inside a planar cavity}
In this section, we assume that the atomic monolayer is placed in
a planar cavity with reflection and transmission coefficients $r_{c}$,
$t_{c}$, respectively~{[}see Fig.~4 of the main manuscript{]}.
By using the transfer matrix approach, we can obtain the reflection
and transmission coefficients of the atomic monolayer inside the planar
cavity. The transfer matrix of the atomic monolayer/bilayer metasurface
can be written as
\begin{eqnarray}
M_{AL} & = & \frac{1}{t_{AL}}\left[\begin{array}{cc}
t_{AL}^{2}-r_{AL}^{2} & r_{AL}\\
-r_{AL} & 1
\end{array}\right],
\end{eqnarray}
where $r_{AL}$ and $t_{AL}$ are the reflected and transmitted coefficients
of the atomic metasurface, respectively {[}see Eqs.~\ref{eq:r_ML}
and \ref{eq:TR_BL-SM}{]}. For the We assumed that the transmission
and reflection coefficients are identical for forward and backward
directions. The nonradiative losses is zero, therefore, the conservation
of energy yields to following formulas
\begin{eqnarray}
\left|r_{AL}\right|^{2}+\left|t_{AL}\right|^{2} & = & 1,\nonumber \\
\frac{t_{AL}}{t_{AL}^{*}} & = & -\frac{r_{AL}}{r_{AL}^{*}},
\end{eqnarray}
thus the following relation holds for the transfer matrix, i.e. $\mathrm{det}\,\left(M\right)=1$,
and
\begin{eqnarray}
M_{AL} & = & \left[\begin{array}{cc}
\frac{1}{t_{AL}^{*}} & \frac{r_{AL}}{t_{AL}}\\
-\frac{r_{AL}^{*}}{t_{AL}^{*}} & \frac{1}{t_{AL}}
\end{array}\right].
\end{eqnarray}
The transmission coefficient from an atomic monolayer/bilayer inside
a cavity read as
\begin{equation}
t=\frac{-t_{M}^{2}\left|t_{AL}\right|^{2}}{r_{M}^{2}t_{AL}e^{ikL_{\mathrm{c}}}+2r_{M}r_{AL}t_{AL}^{*}\mathrm{cos}\left(2kD\right)-t_{AL}^{*}e^{-ikL_{\mathrm{c}}}},\label{eq:t_Cavity_AL}
\end{equation}
The reflection and transmission coefficients of an atomic monolayer
metasurface (AMM) can be written as {[}see Eq.~\ref{eq:r_monolayer}{]}
\begin{equation}
r_{AL}=\frac{-i\Gamma_{AL}/2}{\delta_{AL}+i\Gamma_{AL}/2},\,\,\,\,\frac{\Gamma_{AL}}{2}=\frac{k\alpha_{0}\Gamma_{0}}{4\Lambda^{2}},\,\,\,\,\delta_{AL}=\omega-\omega_{a}-\Delta,\label{eq:r_AL}
\end{equation}
Now, by using $t_{AL}=1+r_{AL}=\frac{\delta_{AL}}{\delta_{AL}+i\Gamma_{AL}/2}$
and substituting Eq.~\ref{eq:r_AL} into Eq.~\ref{eq:t_Cavity_AL},
we obtain
\begin{equation}
t_{\mathrm{AMM}}=\frac{t_{M}^{2}\delta_{AL}e^{ikL_{\mathrm{c}}}}{\delta_{AL}\left(1-e^{2ikL_{\mathrm{c}}}r_{M}^{2}\right)+i\frac{\Gamma_{a}}{2}\left[1+2r_{M}e^{ikL_{\mathrm{c}}}\mathrm{cos}\left(2kD\right)+r_{M}^{2}e^{2ikL_{\mathrm{c}}}\right]}.
\end{equation}
Note that the atomic metasurface is placed at a distance $D$ (see
Fig.4 of the main manuscript). The planar cavity consisting of two
mirrors separated by $L_{c}.$ A similar expression can be obtained
for an atomic bilayer metasurface (ABM) in a cavity
\begin{equation}
t_{\mathrm{ABM}}=-\frac{r_{AL}t_{M}^{2}\left|r_{AL}+1\right|^{2}e^{ikL_{\mathrm{c}}}}{-\left|r_{AL}\right|^{2}-r_{AL}+r_{M}^{2}(r_{AL}^{2}+r_{AL})e^{2ikL_{\mathrm{c}}}-2ie^{ikL_{\mathrm{c}}}r_{21}r_{AL}\left[\left|r_{AL}\right|^{2}\mathrm{sin}\left(2kD\right)+\mathrm{Im}\left(r_{AL}^{*}e^{2ikD}\right)\right]}.
\end{equation}
Note that $r_{AL}$ and $t_{AL}$ for the AMM and ABM can be obtained
from Eq.~\ref{eq:r_monolayer} and Eq.~\ref{eq:TR_BL-SM}, respectively.
\subsection{Scattered fields from an atomic bilayer metasurface using a Gaussian
beam excitation}
For an experimental realization limited number of atoms and a Gaussian
beam excitation would be necessary. The Gaussian beam is defined as
\begin{eqnarray}
\mathbf{E}_{\mathrm{inc}}\left(x,y,z\right) & = & \mathbf{e}_{x}E_{0}\frac{w_{0}}{w\left(z\right)}e^{ikz}e^{-i\varphi\left(z\right)}e^{-\frac{x^{2}+y^{2}}{w^{2}\left(z\right)}}e^{ik\frac{x^{2}+y^{2}}{2R\left(z\right)}},\label{eq:GaussianBeam}
\end{eqnarray}
the beam parameters are defined as
\begin{eqnarray}
w\left(z\right) & = & w_{0}\sqrt{1+\left(\frac{z}{z_{R}}\right)^{2}},\,\,\,\,z_{R}=\frac{\pi w_{0}^{2}}{\lambda},\nonumber \\
\varphi\left(z\right) & = & \mathrm{arctang}\left(\frac{z}{z_{R}}\right),\nonumber \\
R\left(z\right) & = & z\left[1+\left(\frac{z_{R}}{z}\right)^{2}\right],
\end{eqnarray}
where $w_{0}$ is the beam waist at its focal point, $R\left(z\right)$
is the radius of curvature of the beam at $z$, and $\varphi\left(z\right)$
is the Gouy phase at $z$. The Gaussian beam is propagating along
the $z$ direction. In our numerical calculations, we assumed the
cross section of the Gaussian beam is smaller than the area of the
finite atomic layer. The parameters of Gaussian beam is given in the
caption of Fig.~\ref{fig:GaussianResults}. The incident and scattered
electric fields for an electric and a magnetic mirror using such a
Gaussian beam are demonstrated in Fig.~\ref{fig:GaussianResults}.
For the symmetric mode ($z>0$), the scattered field is opposite in
sign with respect to the incident (both real and imaginary). Thus,
the transmitted field which is the sum of the incident and scattered
electric fields is zero ($E_{t}=E_{\mathrm{inc}}+E_{\mathrm{sca}}=0$)
{[}Fig.~\ref{fig:GaussianResults}, see the blue box{]}. However,
for ($z<0$), only the \textit{real} part of the scattered field is
opposite in sign with respect to the \textit{real} part of the incident.
Thus it acts as a electric mirror in the symmetric mode analogous
to a perfect electric conductor. For the antisymmetric mode ($z>0$)
similar to the symmetric mode ($z>0$), the scattered field is opposite
in sign with respect to the incident (both real and imaginary). Thus,
the transmitted field which is the sum of the incident and scattered
electric fields is zero ($E_{t}=E_{\mathrm{inc}}+E_{\mathrm{sca}}=0$)
{[}Fig.~\ref{fig:GaussianResults}, see the red box{]}. However,
for ($z<0$), the \textit{imaginary} part of the scattered field is
opposite in sign with respect to the \textit{imaginary} part of the
incident. Thus it acts as a magnetic mirror analogous to a perfect
magnetic conductor. Using Gaussian beam, we also calculated transmission
and reflection for limited number of atoms {[}$15\times15\times2$
atoms{]}. The results for the two illuminations (i.e. Gaussian and
plane wave) are in a excellent agreement~{[}see the main manuscript
Fig.~3{]}.
\begin{figure}
\begin{centering}
\includegraphics[width=14cm]{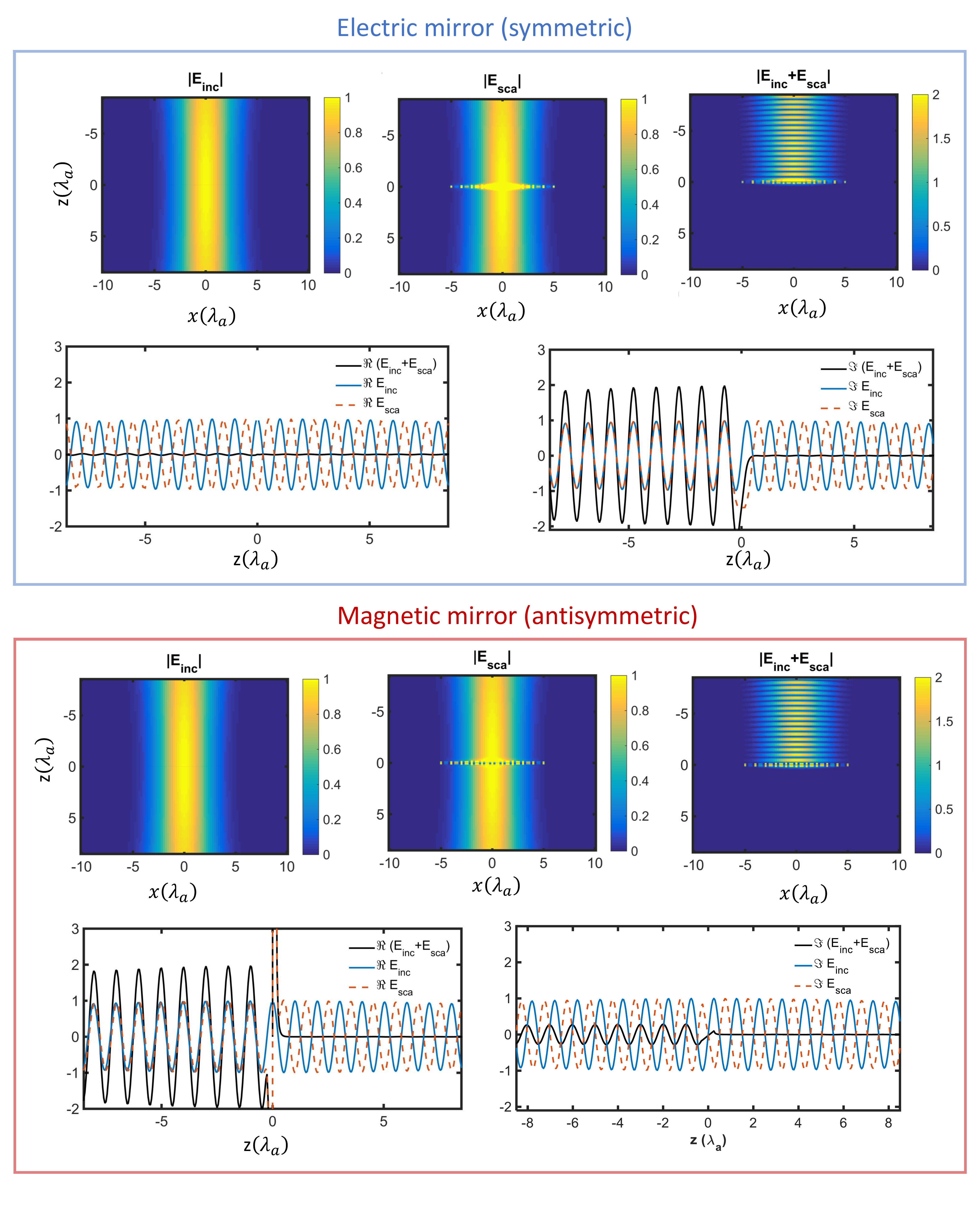}
\par\end{centering}
\caption{Atomic bilayer composed of $15\times15\times2$ atoms when illuminated
by a Gaussian beam at the symmetric (electric mirror) and antisymmetric
(magnetic mirror) modes. We considered a normal incidence where the
cross section of the Gaussian beam is at the center of the atomic
bilayer, z = 0 with $E_{0}=1$, and $W_{0}=\lambda_{a}$. $E_{\mathrm{inc}}$
and $E_{\mathrm{sca}}$ are the incident and scattered electric fields,
respectively. It can be seen that the scattering and incident fields
destructively interfere at the forward direction. \label{fig:GaussianResults}}
\end{figure}

\section{Green functions and scattered fields}

In this section, we define the Green function which is used to calculation
the interaction constant for the atomic dimer and atomic array. The
electric and magnetic dyadic Green functions in free space, respectively,
read~\cite{tai1994dyadic}
\begin{equation}
\bar{\bar{{\bf G}}}_{E}\left(\mathbf{r},\mathbf{r}^{\prime}\right)=\left({\bf \bar{\bar{I}}}+\frac{1}{k^{2}}\nabla\nabla\right)G_{0}\left(\mathbf{r},\mathbf{r}^{\prime}\right),
\end{equation}
and
\begin{equation}
\bar{\bar{{\bf G}}}_{M}\left(\mathbf{r},\mathbf{r}^{\prime}\right)=\nabla\times\left[{\bf \bar{\bar{I}}}G_{0}\left(\mathbf{r},\mathbf{r}^{\prime}\right)\right]=\nabla G_{0}\left(\mathbf{r},\mathbf{r}^{\prime}\right)\times{\bf \bar{\bar{I}}},\label{eq:magnetic dyadic}
\end{equation}
where, $G_{0}\left(\mathbf{r},\mathbf{r}^{\prime}\right)={\rm e^{{\it ik\vert\mathbf{r}-\mathbf{r}^{\prime}\vert}}/\left(4\pi\vert\mathbf{r}-\mathbf{r}^{\prime}\vert\right)}$
is the scalar free space Green function and the identity $\nabla\times\left(G_{0}{\bf \bar{\bar{I}}}\right)=G_{0}\nabla\times{\bf \bar{\bar{I}}}+\nabla G_{0}\times{\bf \bar{\bar{I}}}$
(where ${\bf \bar{\bar{I}}}$ is the identity dyadic) is used in Eq.
(\ref{eq:magnetic dyadic}). The electric and magnetic fields in term
of electric vector potential ${\bf A}$ read ${\bf E}=i\omega\left({\bf A}+\frac{1}{k^{2}}\nabla\nabla\cdot{\bf A}\right)$
and ${\bf H}=\frac{1}{\mu_{0}}\left(\nabla\times{\bf A}\right)$,
respectively ($\nabla$ is taken over variable$\mathbf{r}$). For
an electric dipole ${\bf p}$, the electric vector potential ${\bf A}$
reads~\cite{Jackson1999}
\begin{equation}
{\bf A}\left(\mathbf{r}\right)=-i\omega\mu_{0}\frac{e^{{\it ik\vert\mathbf{r}-\mathbf{r}^{\prime}\vert}}}{4\pi\vert\mathbf{r}-\mathbf{r}^{\prime}\vert}{\bf p}\left(\mathbf{r}^{\prime}\right)=-i\omega\mu_{0}G_{0}\left(\mathbf{r},\mathbf{r}^{\prime}\right){\bf p}\left(\mathbf{r}^{\prime}\right).
\end{equation}
Therefore, the electric and magnetic fields ${\bf E_{p}}$ and ${\bf H_{p}}$
of an electric dipole moment, respectively, read
\begin{equation}
{\bf E_{p}}\left(\mathbf{r}\right)=\omega^{2}\mu_{0}\bar{\bar{{\bf G}}}_{E}\left(\mathbf{r},\mathbf{r}^{\prime}\right)\cdot{\bf p}\left(\mathbf{r}^{\prime}\right),\label{e_filed_p}
\end{equation}
\begin{equation}
{\bf H_{p}\left(\mathbf{r}\right)}=-i\omega\bar{\bar{{\bf G}}}_{M}\left(\mathbf{r},\mathbf{r}^{\prime}\right)\cdot{\bf p}\left(\mathbf{r}^{\prime}\right).\label{H_filed_p}
\end{equation}
For a magnetic dipole ${\bf m}$, the electric and magnetic fields
${\bf E_{m}}$ and ${\bf H_{m}}$, respectively, read
\begin{equation}
{\bf E_{m}}\left(\mathbf{r}\right)=i\omega Z_{0}\bar{\bar{{\bf G}}}_{M}\left(\mathbf{r},\mathbf{r}^{\prime}\right)\cdot\frac{{\bf m}\left(\mathbf{r}^{\prime}\right)}{c},\label{e_filed_m}
\end{equation}
\begin{equation}
{\bf H_{m}\left(\mathbf{r}\right)}=\omega^{2}\epsilon_{0}Z_{0}\bar{\bar{{\bf G}}}_{E}\left(\mathbf{r},\mathbf{r}^{\prime}\right)\cdot\frac{{\bf m}\left(\mathbf{r}^{\prime}\right)}{c},\label{h_filed_m}
\end{equation}
since the electric vector potential for the dipole ${\bf m}$ reads~\cite{Jackson1999}
\begin{equation}
{\bf A}\left(\mathbf{r}\right)=Z_{0}\nabla G_{0}\left(\mathbf{r},\mathbf{r}^{\prime}\right)\times\frac{{\bf m}\left(\mathbf{r}^{\prime}\right)}{c}.\label{eq:vector_pt_m}
\end{equation}
Note that we have used~\cite{Jackson1999}
\begin{equation}
{\bf A}\left(\mathbf{r}\right)=i\omega\mu_{0}\frac{e^{{\it ik\vert\mathbf{r}-\mathbf{r}^{\prime}\vert}}}{4\pi\vert\mathbf{r}-\mathbf{r}^{\prime}\vert}\left(1-\frac{1}{ik\vert\mathbf{r}-\mathbf{r}^{\prime}\vert}\right){\bf \left({\bf n}\times\frac{{\bf m}\left(\mathbf{r}^{\prime}\right)}{{\it c}}\right)},
\end{equation}
and
\begin{equation}
\nabla G_{0}\left(\mathbf{r},\mathbf{r}^{\prime}\right)=ikG_{0}\left(\mathbf{r},\mathbf{r}^{\prime}\right)\left(1-\frac{1}{ik\vert\mathbf{r}-\mathbf{r}^{\prime}\vert}\right){\bf \frac{\left(\mathbf{r}-\mathbf{r}^{\prime}\right)}{\vert\mathbf{r}-\mathbf{r}^{\prime}\vert}},\,\,\,\,{\bf n}=\frac{\left(\mathbf{r}-\mathbf{r}^{\prime}\right)}{\vert\mathbf{r}-\mathbf{r}^{\prime}\vert},
\end{equation}
in Eq. (\ref{eq:vector_pt_m}). Also note that we have used identity
${\bf a\times{\bf b}}={\bf a}\times\left(\bar{\bar{{\bf I}}}\cdot{\bf b}\right)=\left({\bf a}\times\bar{\bar{{\bf I}}}\right)\cdot{\bf b}$
in obtaining Eq.(\ref{e_filed_m}). Finally, the fields created by
an electric and a magnetic dipole from Eqs. (\ref{e_filed_p})-(\ref{e_filed_m})
read
\begin{eqnarray}
\mathbf{E}\left(\mathbf{r}\right) & = & {\bf E_{p}}\left(\mathbf{r}\right)+{\bf E_{m}\left(\mathbf{r}\right)}=\mathbf{\bar{\bar{G}}}_{EE}\left(\mathbf{r},\mathbf{r}^{\prime}\right)\cdot\mathbf{p}\left(\mathbf{r}^{\prime}\right)+{\bf \bar{\bar{G}}}_{EM}\left(\mathbf{r},\mathbf{r}^{\prime}\right)\cdot\frac{\mathbf{m}\left(\mathbf{r}^{\prime}\right)}{c},\nonumber \\
Z_{0}\mathbf{H}\left(\mathbf{r}\right) & = & Z_{0}\left[{\bf H_{p}}\left(\mathbf{r}\right)+{\bf H_{m}}\left(\mathbf{r}\right)\right]={\bf \bar{\bar{G}}}_{ME}\left(\mathbf{r},\mathbf{r}^{\prime}\right)\cdot\mathbf{p}\left(\mathbf{r}^{\prime}\right)+\mathbf{\bar{\bar{G}}}_{MM}\left(\mathbf{r},\mathbf{r}^{\prime}\right)\cdot\frac{\mathbf{m}\left(\mathbf{r}^{\prime}\right)}{c},\label{eq:EH_GreenFun}
\end{eqnarray}
where,

\begin{eqnarray}
{\bf \bar{\bar{G}}}_{EE}\left(\mathbf{r},\mathbf{r}^{\prime}\right) & = & \omega^{2}\mu_{0}\bar{\bar{{\bf G}}}_{E}\left(\mathbf{r},\mathbf{r}^{\prime}\right),\,\,\,{\bf \bar{\bar{G}}}_{EM}\left(\mathbf{r},\mathbf{r}^{\prime}\right)=i\omega Z_{0}\bar{\bar{{\bf G}}}_{M}\left(\mathbf{r},\mathbf{r}^{\prime}\right),\nonumber \\
\mathbf{\bar{\bar{G}}}_{ME}\left(\mathbf{r},\mathbf{r}^{\prime}\right) & = & -{\bf \bar{\bar{G}}}_{EM}\left(\mathbf{r},\mathbf{r}^{\prime}\right),\,\,\,\mathbf{\bar{\bar{G}}}_{MM}\left(\mathbf{r},\mathbf{r}^{\prime}\right)={\bf \bar{\bar{G}}}_{EE}\left(\mathbf{r},\mathbf{r}^{\prime}\right),
\end{eqnarray}
and can be written as

\begin{equation}
{\bf \bar{\bar{G}}}_{EE}\left(\mathbf{r},\mathbf{r}^{\prime}\right)=\mathbf{\bar{\bar{G}}}_{MM}\left(\mathbf{r},\mathbf{r}^{\prime}\right)=\frac{3}{2\alpha_{0}\epsilon_{0}}e^{i\zeta}\left[\left(\frac{1}{\zeta}-\frac{1}{\zeta^{3}}+\frac{i}{\zeta^{2}}\right)\bar{\bar{{\bf I}}}+\left(-\frac{1}{\zeta}+\frac{3}{\zeta^{3}}-\frac{3i}{\zeta^{2}}\right)\mathbf{\mathbf{n}n}\right],
\end{equation}

\begin{equation}
{\bf \bar{\bar{G}}}_{EM}\left(\mathbf{r},\mathbf{r}^{\prime}\right)=-\mathbf{\bar{\bar{G}}}_{ME}\left(\mathbf{r},\mathbf{r}^{\prime}\right)=-\frac{3}{2\alpha_{0}\epsilon_{0}}e^{i\zeta}\left(\frac{1}{\zeta}-\frac{1}{i\zeta^{2}}\right)\mathbf{n}\times\bar{\bar{{\bf I}}},
\end{equation}
where $\mathbf{n}=\frac{\mathbf{r}-\mathbf{r}^{\prime}}{\left|\mathbf{r}-\mathbf{r}^{\prime}\right|}$,
$\alpha_{0}=\frac{6\pi}{k^{3}}$ and $,\zeta=k\left(\mathbf{r}-\mathbf{r}^{\prime}\right)$
and can be also written as

\begin{eqnarray}
G_{EE}^{\alpha\beta}\left(\zeta=k\left|\mathbf{r}-\mathbf{r}^{\prime}\right|\right) & = & \frac{3}{2\alpha_{0}\epsilon_{0}}e^{i\zeta}\left[g_{1}\left(\zeta\right)\delta_{\alpha\beta}+g_{2}\left(\zeta\right)\frac{\zeta_{\alpha}\zeta_{\beta}}{\zeta^{2}}\right],\nonumber \\
g_{1}\left(\zeta\right) & = & \left(\frac{1}{\zeta}-\frac{1}{\zeta^{3}}+\frac{i}{\zeta^{2}}\right),\nonumber \\
g_{2}\left(\zeta\right) & = & \left(-\frac{1}{\zeta}+\frac{3}{\zeta^{3}}-\frac{3i}{\zeta^{2}}\right),
\end{eqnarray}

Note that Eq.~\ref{eq:EH_GreenFun} can be also written as

\begin{eqnarray}
\mathbf{E}\left(\mathbf{r}\right) & = & {\bf E_{p}}\left(\mathbf{r}\right)+{\bf E_{m}\left(\mathbf{r}\right)}=\mathbf{\bar{\bar{G}}}_{EE}\left(\mathbf{r},\mathbf{r}^{\prime}\right)\cdot\mathbf{p}\left(\mathbf{r}^{\prime}\right)+g_{EM}\mathbf{n}\times\frac{\mathbf{m}\left(\mathbf{r}^{\prime}\right)}{c},\nonumber \\
Z_{0}\mathbf{H}\left(\mathbf{r}\right) & = & Z_{0}\left[{\bf H_{p}}\left(\mathbf{r}\right)+{\bf H_{m}}\left(\mathbf{r}\right)\right]=g_{ME}\mathbf{n}\times\mathbf{p}\left(\mathbf{r}^{\prime}\right)+\mathbf{\bar{\bar{G}}}_{MM}\left(\mathbf{r},\mathbf{r}^{\prime}\right)\cdot\frac{\mathbf{m}\left(\mathbf{r}^{\prime}\right)}{c},\label{eq:EH_GreenFunFinal}
\end{eqnarray}
where $g_{ME}=-g_{EM}=\frac{3}{2\alpha_{0}\epsilon_{0}}e^{i\zeta}\left(\frac{1}{\zeta}-\frac{1}{i\zeta^{2}}\right)$.

\section{Enhancing the decay rate of a magnetic emitter}

\subsection{Coupled dipole theory}
Let us consider N atoms with electric dipole transition moments in
free space. The self-consistent equation for the induced dipole moments
of $i$th atom placed at $\mathbf{r}=\mathbf{r}_{i}$ read~\cite{foldy1945,Mulholland:94,lagendijk1996,Alaee:2017Review}
\begin{eqnarray}
\mathbf{p}\left(\mathbf{r}_{i}\right) & = & \epsilon_{0}\alpha_{i}\left[\mathbf{E}_{\mathrm{inc}}\left(\mathbf{r}_{i}\right)+\underset{i\neq j}{\sum}\mathbf{E}_{\mathrm{sca}}\left(\left|\mathbf{r}_{j}-\mathbf{r}_{i}\right|\right)\right],\label{eq:CDT}
\end{eqnarray}
where $\mathbf{E}_{\mathrm{inc}}\left(\mathbf{r}_{i}\right)$ is the
incident field at the atom position, $\alpha_{i}$ is the atomic polarizability
and $\underset{i\neq j}{\sum}\mathbf{E}_{\mathrm{sca}}\left(\left|\mathbf{r}_{j}-\mathbf{r}_{i}\right|\right)$
are the interaction fields created by the all atoms at $\mathbf{r}=\mathbf{r}_{i}$.
Using Eq.~\ref{eq:CDT}, we can obtain
\begin{eqnarray}
\underset{i\neq j}{\sum}\mathbf{E}_{\mathrm{sca}}\left(\left|\mathbf{r}_{j}-\mathbf{r}_{i}\right|\right) & = & \mathbf{p}\left(\mathbf{r}_{i}\right)/\epsilon_{0}\alpha_{i}-\mathbf{E}_{\mathrm{inc}}\left(\mathbf{r}_{i}\right),
\end{eqnarray}
Thus, the total field at $\mathbf{r}=\mathbf{r}_{i}$, can be calculated
by using
\begin{eqnarray}
\mathbf{E}_{\mathrm{tot}}\left(\mathbf{r}_{i}\right) & = & \mathbf{E}_{\mathrm{inc}}\left(\mathbf{r}_{i}\right)+\underset{i}{\sum}\mathbf{E}_{\mathrm{sca}}\left(\left|\mathbf{r}_{j}-\mathbf{r}_{i}\right|\right),\nonumber \\
 & = & \mathbf{E}_{\mathrm{inc}}\left(\mathbf{r}_{i}\right)+\mathbf{E}_{\mathrm{sca}}\left(\left|\mathbf{r}_{i}-\mathbf{r}_{j}\right|=0\right)+\underset{i\neq j}{\sum}\mathbf{E}_{\mathrm{sca}}\left(\left|\mathbf{r}_{j}-\mathbf{r}_{i}\right|\right),\nonumber \\
 & = & \mathbf{E}_{\mathrm{inc}}\left(\mathbf{r}_{i}\right)+\mathbf{\bar{\bar{G}}}_{EE}\left(\left|\mathbf{r}_{i}-\mathbf{r}_{j}\right|=0\right)\cdot\mathbf{p}\left(\mathbf{r}_{i}\right)+\frac{\mathbf{p}\left(\mathbf{r}_{i}\right)}{\epsilon_{0}\alpha_{i}}-\mathbf{E}_{\mathrm{inc}}\left(\mathbf{r}_{i}\right),\nonumber \\
 & = & \left[\mathbf{\bar{\bar{G}}}_{EE}\left(\left|\mathbf{r}_{i}-\mathbf{r}_{j}\right|=0\right)\cdot\mathbf{p}\left(\mathbf{r}_{i}\right)+\frac{1}{\epsilon_{0}\alpha_{i}}\mathbf{p}\left(\mathbf{r}_{i}\right)\right],\label{eq:Etot}
\end{eqnarray}
where we used $\mathbf{E}_{\mathrm{sca}}\left(\left|\mathbf{r}_{i}-\mathbf{r}_{j}\right|=0\right)=\mathbf{\bar{\bar{G}}}_{EE}\left(\left|\mathbf{r}_{i}-\mathbf{r}_{j}\right|=0\right)\cdot\mathbf{p}\left(\mathbf{r}_{i}\right)$.
Now by using the total field, we can compute the absorbed power
\begin{eqnarray}
P_{\mathrm{abs}} & = & -\frac{\omega}{2}\mathrm{Im}\left[\underset{i}{\sum}\mathbf{p}^{*}\left(\mathbf{r}_{i}\right)\cdot\mathbf{E}_{\mathrm{tot}}\left(\mathbf{r}_{i}\right)\right],\nonumber \\
 & = & -\frac{\omega}{2}\mathrm{Im}\left\{ \underset{i}{\sum}\mathbf{p}^{*}\left(\mathbf{r}_{i}\right)\cdot\left[\mathbf{\bar{\bar{G}}}_{EE}\left(\left|\mathbf{r}_{i}-\mathbf{r}_{j}\right|=0\right)\cdot\mathbf{p}\left(\mathbf{r}_{i}\right)+\frac{1}{\epsilon_{0}\alpha_{i}}\mathbf{p}\left(\mathbf{r}_{i}\right)\right]\right\} ,\nonumber \\
 & = & -\frac{\omega}{2\epsilon_{0}}\underset{i}{\sum}\left|\mathbf{p}\left(\mathbf{r}_{i}\right)\right|^{2}\left(\frac{1}{\alpha_{0}}+\mathrm{Im}\left[\frac{1}{\alpha_{i}}\right]\right),\label{eq:P_abs}
\end{eqnarray}Note that $\mathrm{Im}\left[\mathbf{\bar{\bar{G}}}_{EE}\left(\left|\mathbf{r}_{i}-\mathbf{r}_{j}\right|=0\right)\right]=\frac{1}{\epsilon_{0}\alpha_{0}}\bar{\bar{{\bf I}}}$~\cite{lagendijk1996,Lagendijk1998,Alaee:2017Review}
and for lossless dipoles $\mathrm{Im}\left[\frac{1}{\alpha_{i}}\right]=-\frac{1}{\alpha_{0}}$,
therefore $P_{\mathrm{abs}}=0$. The scattered power can be calculated
as
\begin{eqnarray}
P_{\mathrm{sca}} & = & \frac{\omega}{2}\mathrm{Im}\left[\underset{i}{\sum}\mathbf{p}^{*}\left(\mathbf{r}_{i}\right)\cdot\underset{j}{\sum}\mathbf{E}_{\mathrm{sca}}\left(\left|\mathbf{r}_{j}-\mathbf{r}_{i}\right|\right)\right],\nonumber \\
 & = & \frac{\omega}{2}\mathrm{Im}\left\{ \underset{i}{\sum}\mathbf{p}^{*}\left(\mathbf{r}_{i}\right)\cdot\left[\mathbf{\bar{\bar{G}}}_{EE}\left(\left|\mathbf{r}_{i}-\mathbf{r}_{j}\right|=0\right)\cdot\mathbf{p}\left(\mathbf{r}_{i}\right)+\mathbf{p}\left(\mathbf{r}_{i}\right)/\epsilon_{0}\alpha_{i}-\mathbf{E}_{\mathrm{inc}}\left(\mathbf{r}_{i}\right)\right]\right\} \nonumber \\
 & = & \frac{\omega}{2}\mathrm{Im}\left\{ \underset{i}{\sum}\mathbf{p}_{i}^{*}\left(\mathbf{r}_{i}\right)\cdot\left[\frac{1}{\epsilon_{0}\alpha_{0}}\mathbf{p}_{i}\left(\mathbf{r}_{i}\right)+\frac{1}{\epsilon_{0}\alpha_{i}}\mathbf{p}_{i}\left(\mathbf{r}_{i}\right)-\mathbf{E}_{\mathrm{inc}}\left(\mathbf{r}_{i}\right)\right]\right\} \nonumber \\
 & = & \frac{\omega}{2}\underset{i}{\sum}\left|\mathbf{p}\left(\mathbf{r}_{i}\right)\right|^{2}\left(\frac{1}{\epsilon_{0}\alpha_{0}}+\mathrm{Im}\left[\frac{1}{\epsilon_{0}\alpha_{i}}\right]\right)-\frac{\omega}{2}\underset{i}{\sum}\mathrm{Im}\left[\mathbf{p}^{*}\left(\mathbf{r}_{i}\right)\cdot\mathbf{E}_{\mathrm{inc}}\left(\mathbf{r}_{i}\right)\right],\nonumber \\
 & = & -P_{\mathrm{abs}}-\frac{\omega}{2}\underset{i}{\sum}\mathrm{Im}\left[\mathbf{p}^{*}\left(\mathbf{r}_{i}\right)\cdot\mathbf{E}_{\mathrm{inc}}\left(\mathbf{r}_{i}\right)\right],\label{eq:Psca}
\end{eqnarray}
now by using power conservation,  the extracted power can be defined
as

\begin{eqnarray}
P_{\mathrm{ext}} & = & -\frac{\omega}{2}\underset{i}{\sum}\mathrm{Im}\left[\mathbf{p}_{i}^{*}\left(\mathbf{r}_{i}\right)\cdot\mathbf{E}_{\mathrm{inc}}\left(\mathbf{r}_{i}\right)\right]=P_{\mathrm{sca}}+P_{\mathrm{abs}}.\label{eq:Pext}
\end{eqnarray}

In conclusion, the absorbed, scattered, and extracted powers read
as

\begin{equation}
P_{\mathrm{abs}}=-\frac{\omega}{2\epsilon_{0}}\underset{i}{\sum}\left|\mathbf{p}\left(\mathbf{r}_{i}\right)\right|^{2}\left(\frac{1}{\alpha_{0}}+\mathrm{Im}\left[\frac{1}{\alpha_{i}}\right]\right),
\end{equation}

\begin{equation}
P_{\mathrm{sca}}=\frac{\omega}{2\epsilon_{0}}\underset{i}{\sum}\left|\mathbf{p}\left(\mathbf{r}_{i}\right)\right|^{2}\left(\frac{1}{\alpha_{0}}+\mathrm{Im}\left[\frac{1}{\alpha_{i}}\right]\right)-\frac{\omega}{2}\underset{i}{\sum}\mathrm{Im}\left[\mathbf{p}^{*}\left(\mathbf{r}_{i}\right)\cdot\mathbf{E}_{\mathrm{inc}}\left(\mathbf{r}_{i}\right)\right],
\end{equation}

\begin{equation}
P_{\mathrm{ext}}=-\frac{\omega}{2}\underset{i}{\sum}\mathrm{Im}\left[\mathbf{p}^{*}\left(\mathbf{r}_{i}\right)\cdot\mathbf{E}_{\mathrm{inc}}\left(\mathbf{r}_{i}\right)\right]=\frac{\omega}{2}\underset{i}{\sum}\mathrm{Im}\left[\mathbf{p}\left(\mathbf{r}_{i}\right)\cdot\mathbf{E}^{*}_{\mathrm{inc}}\left(\mathbf{r}_{i}\right)\right].
\end{equation}

Using duality ($\mathbf{p}\leftrightarrow\frac{\mathbf{m}}{c}$ and
$\mathbf{E}\leftrightarrow\frac{\mathbf{H}}{Z_{0}}$~see Ref.~\cite{Jackson1999}),
similar results can be obtained for a magnetic dipole moment, i.e. 

\begin{equation}
P_{\mathrm{abs}}=-\frac{\omega}{2\epsilon_{0}}\underset{i}{\sum}\left|\frac{\mathbf{m}\left(\mathbf{r}_{i}\right)}{c}\right|^{2}\left(\frac{1}{\alpha_{0}}+\mathrm{Im}\left[\frac{1}{\alpha_{i}}\right]\right),
\end{equation}

\begin{equation}
P_{\mathrm{sca}}=\frac{\omega}{2\epsilon_{0}}\underset{i}{\sum}\left|\frac{\mathbf{m}\left(\mathbf{r}_{i}\right)}{c}\right|^{2}\left(\frac{1}{\alpha_{0}}+\mathrm{Im}\left[\frac{1}{\alpha_{i}}\right]\right)-\frac{\omega}{2}\underset{i}{\sum}\mathrm{Im}\left[\frac{\mathbf{m}^{*}\left(\mathbf{r}_{i}\right)}{c}\cdot\frac{\mathbf{H}_{\mathrm{inc}}\left(\mathbf{r}_{i}\right)}{Z_{0}}\right],
\end{equation}

\begin{equation}
P_{\mathrm{ext}}=-\frac{\omega}{2}\underset{i}{\sum}\mathrm{Im}\left[\frac{\mathbf{m}^{*}\left(\mathbf{r}_{i}\right)}{c}\cdot\frac{\mathbf{H}_{\mathrm{inc}}\left(\mathbf{r}_{i}\right)}{Z_{0}}\right]=\frac{\omega}{2}\underset{i}{\sum}\mathrm{Im}\left[\frac{\mathbf{m}\left(\mathbf{r}_{i}\right)}{c}\cdot\frac{\mathbf{H}^{*}_{\mathrm{inc}}\left(\mathbf{r}_{i}\right)}{Z_{0}}\right].
\end{equation}

In the next subsection, we used above expressions to obtain the radiated
power of a test magnetic dipole emitter.

\subsection{Emission rate enhancement of a magnetic emitter}
The power radiated by a test magnetic dipole emitter, i.e. $\bm{\mu}_{t}=\mu_{t}\mathbf{n}_{\mu}$
placed at $\mathbf{r}_{0}$ is given by~\cite{Jackson1999}
\begin{eqnarray}
P_{\mathrm{rad}}^{\mathrm{fs}} & = & \frac{ck^{4}}{12\pi\epsilon_{0}}\left|\frac{\bm{\mu}_{t}}{c}\right|^{2}=\frac{1}{2}\frac{\omega}{\alpha_{0}\epsilon_{0}}\left|\frac{\bm{\mu}_{t}}{c}\right|^{2},\label{eq:P_rad_MD_fs-1}
\end{eqnarray}
where $\mathbf{n}_{\mu}$ is the unit vector in the direction of the
dipole moment. The power radiated by a magnetic dipole emitter $\bm{\mu}_{t}=\bm{\mu}_{t}\left(\mathbf{r}_{0}\right)$
when placed close to a antenna consist of $N$ atoms at position $\mathbf{r}_{i}$
with \textit{only} electric dipole moment $\mathbf{p}_{i}$, read
as
\begin{eqnarray}
P_{\mathrm{rad}}^{\mathrm{ant}} & = & \frac{\omega}{2}\mathrm{Im}\left[\frac{\bm{\mu}_{t}}{c}^{*}\cdot Z_{0}\mathbf{H}_{\mathrm{local}}\left(\mathbf{r}_{0}\right)\right],\label{eq:P_rad_MD_ant}\\
 & = & \frac{\omega}{2}\left|\frac{\bm{\mu}_{t}}{c}\right|^{2}\mathrm{Im}\left[\mathbf{n}_{\mu}^{T}\mathbf{G}_{\mathrm{tot}}\left(\mathbf{r}_{0},\mathbf{r}_{0}\right)\mathbf{n}_{\mu}\right],\end{eqnarray}
where $\mathbf{G}_{\mathrm{tot}}\left(\mathbf{r}_{0},\mathbf{r}_{0}\right)$
is the total Green function at the position of the emitter, and the
$\mathbf{H}\left(\mathbf{r}_{0}\right)$ is the electric field at
the dipole position $\mathbf{r}_{0}$ can be obtained by
\begin{eqnarray}
Z_{0}\mathbf{H}_{\mathrm{local}}\left(\mathbf{r}_{0}\right) & = & \mathbf{\bar{\bar{G}}}_{MM}\left(\mathbf{r}_{0},\mathbf{r}_{0}\right)\cdot\frac{\bm{\mu}_{t}}{c}+{\sum_{i=1}^N}g_{ME}\left(\mathbf{r}_{0},\mathbf{r}_{i}\right)\left[\mathbf{n}_{r_{0}r_{i}}\times\mathbf{p}\left(\mathbf{r}_{i}\right)\right],\label{eq:AA1}
\end{eqnarray}
where $\mathbf{n}_{r_{0}r_{i}}=\frac{\mathbf{r}_{0}-\mathbf{r}_{i}}{\left|\mathbf{r}_{0}-\mathbf{r}_{i}\right|}$,
$g_{ME}$ is defined in Eq.~\ref{eq:EH_GreenFunFinal} and $\mathbf{p}\left(\mathbf{r}_{i}\right)$
can be calculated by using Eq.~\ref{eq:EH_GreenFunFinal}
\begin{eqnarray}
\mathbf{p}\left(\mathbf{r}_{i}\right) & = & \epsilon_{0}\alpha_{i}\mathbf{E}_{\mathrm{local}}\left(\mathbf{r}_{i}\right)\nonumber \\
 & = & \epsilon_{0}\alpha_{i}\left[\mathbf{E}_{\mu_{t}}\left(\mathbf{r}_{i}\right)+{\sum_{j \neq i}^N}\left[\mathbf{\bar{\bar{G}}}_{EE}\left(\mathbf{r}_{i},\mathbf{r}_{j}\right)\cdot\mathbf{p}\left(\mathbf{r}_{j}\right)\right]\right],
\end{eqnarray}
where $\mathbf{E}_{\mu_{t}}\left(\mathbf{r}_{i}\right)\equiv g_{EM}\left(\mathbf{r}_{i},\mathbf{r}_{0}\right)\left(\mathbf{n}_{r_{i}r_{0}}\times\frac{\bm{\mu}_{t}}{c}\right)$.
Thus it can be written in the following form
\begin{eqnarray}
\mathbf{E}_{\mu_{t}}\left(\mathbf{r}_{i}\right) & = & \frac{1}{\epsilon_{0}\alpha_{i}}\mathbf{p}\left(\mathbf{r}_{i}\right)-{\sum_{j \neq i}^N}\left[\mathbf{\bar{\bar{G}}}_{EE}\left(\mathbf{r}_{i},\mathbf{r}_{j}\right)\cdot\mathbf{p}\left(\mathbf{r}_{j}\right)\right],
\end{eqnarray}
and we can write
\begin{eqnarray}
\left[\overline{\mathbf{p}}\right]_{3N\times1} & = & \left[A\right]_{3N\times3N}\left[\overline{\mathbf{E}}_{m_{0}}\right]_{3N\times1},\label{eq:P_eff}
\end{eqnarray}

where $\left(A^{-1}\right)_{ij}^{\mu\nu}=\frac{1}{\alpha\epsilon_{0}}\delta_{ij}\delta_{\mu\nu}-\left(1-\delta_{ij}\right)G_{EE}^{\mu\nu}\left(\mathbf{r}_{i},\mathbf{r}_{j}\right)$
and $G_{EE}^{\mu\nu}\left(\mathbf{r}_{i},\mathbf{r}_{j}\right)$ is
the shorthand for the $(\mu,\nu)$th matrix element of $\mathbf{\bar{\bar{G}}}_{EE}\left(\mathbf{r}_{i},\mathbf{r}_{j}\right)$
and $(\mu,\nu)\rightarrow\left(x,y,z\right)$. $\overline{\mathbf{E}}_{\mu_{t}}=\left[\begin{array}{cccccc}
g_{EM}\left(\mathbf{r}_{1},\mathbf{r}_{0}\right)\left(\mathbf{n}_{r_{1}r_{0}}\times\frac{\mathbf{m}_{0}}{c}\right), & g_{EM}\left(\mathbf{r}_{2},\mathbf{r}_{0}\right)\left(\mathbf{n}_{r_{2}r_{0}}\times\frac{\mathbf{m}_{0}}{c}\right), &  & \cdots & , & g_{EM}\left(\mathbf{r}_{N},\mathbf{r}_{0}\right)\left(\mathbf{n}_{r_{N}r_{0}}\times\frac{\mathbf{m}_{0}}{c}\right)\end{array}\right]^{T}$ and $\overline{\mathbf{p}}=\left[\begin{array}{cccccc}
\mathbf{p}\left(\mathbf{r}_{1}\right), & \mathbf{p}\left(\mathbf{r}_{2}\right), & \mathbf{p}\left(\mathbf{r}_{1}\right), & \cdots & \mathbf{p}\left(\mathbf{r}_{N-1}\right), & \mathbf{p}\left(\mathbf{r}_{N}\right)\end{array}\right]^{T}$ are $3N\times1$ vectors. $\mathbf{p}\left(\mathbf{r}_{i}\right)=\left[\begin{array}{ccc}
p_{x}\left(\mathbf{r}_{i}\right), & p_{y}\left(\mathbf{r}_{i}\right), & p_{z}\left(\mathbf{r}_{i}\right)\end{array}\right]$ and $i=1,2,3,\ldots,N$. $A$ is the collective polarizability (a
$3N\times3N$ matrix) and read as\\
\begin{equation}
A^{-1}	=	\left[\begin{array}{ccccccc}
\frac{1}{\epsilon_{0}\alpha} & 0 & 0 & -G_{EE}^{xx}\left(\mathbf{r}_{1},\mathbf{r}_{2}\right) & -G_{EE}^{xy}\left(\mathbf{r}_{1},\mathbf{r}_{2}\right) & -G_{EE}^{xz}\left(\mathbf{r}_{1},\mathbf{r}_{2}\right)\\
0 & \frac{1}{\epsilon_{0}\alpha} & 0 & -G_{EE}^{yx}\left(\mathbf{r}_{1},\mathbf{r}_{2}\right) & -G_{EE}^{yy}\left(\mathbf{r}_{1},\mathbf{r}_{2}\right) & -G_{EE}^{yz}\left(\mathbf{r}_{1},\mathbf{r}_{2}\right) & \cdots\\
0 & 0 & \frac{1}{\epsilon_{0}\alpha} & -G_{EE}^{zx}\left(\mathbf{r}_{1},\mathbf{r}_{2}\right) & -G_{EE}^{zy}\left(\mathbf{r}_{1},\mathbf{r}_{2}\right) & -G_{EE}^{zz}\left(\mathbf{r}_{1},\mathbf{r}_{2}\right)\\
-G_{EE}^{xx}\left(\mathbf{r}_{2},\mathbf{r}_{1}\right) & -G_{EE}^{xy}\left(\mathbf{r}_{2},\mathbf{r}_{1}\right) & -G_{EE}^{xz}\left(\mathbf{r}_{2},\mathbf{r}_{1}\right) & \frac{1}{\epsilon_{0}\alpha} & 0 & 0 & 0\\
-G_{EE}^{yx}\left(\mathbf{r}_{2},\mathbf{r}_{1}\right) & -G_{EE}^{yy}\left(\mathbf{r}_{2},\mathbf{r}_{1}\right) & -G_{EE}^{yz}\left(\mathbf{r}_{2},\mathbf{r}_{1}\right) & 0 & \frac{1}{\epsilon_{0}\alpha} & 0 & \cdots\\
-G_{EE}^{zx}\left(\mathbf{r}_{2},\mathbf{r}_{1}\right) & -G_{EE}^{zy}\left(\mathbf{r}_{2},\mathbf{r}_{1}\right) & -G_{EE}^{zz}\left(\mathbf{r}_{2},\mathbf{r}_{1}\right) & 0 & 0 & \frac{1}{\epsilon_{0}\alpha}\\
 & \vdots &  &  & \vdots
\end{array}\right]_{3N\times3N}
\end{equation}
and can be written as
\begin{eqnarray}
\left[\overline{\mathbf{p}}\right]_{i} & =\mathbf{p}\left(\mathbf{r}_{i}\right)= & \left[A\overline{\mathbf{E}}_{\mu_{t}}\right]_{i},\label{eq:P_eff-1}
\end{eqnarray}
Now we can calculate $\mathbf{n}_{r_{0}r_{i}}\times\mathbf{p}\left(\mathbf{r}_{i}\right)$
\begin{eqnarray}
\mathbf{n}_{r_{0}r_{i}}\times\mathbf{p}\left(\mathbf{r}_{i}\right) & = & \mathbf{n}_{r_{0}r_{i}}\times\left[A\overline{\mathbf{E}}_{\mu_{t}}\right]_{i},
\end{eqnarray}
and using Eq.~\ref{eq:AA1} and Eq.~\ref{eq:P_eff-1} we get
\begin{eqnarray}
Z_{0}\mathbf{H}_{\mathrm{local}}\left(\mathbf{r}_{0}\right) & = & \mathbf{\bar{\bar{G}}}_{MM}\left(\mathbf{r}_{0},\mathbf{r}_{0}\right)\cdot\frac{\bm{\mu}_{t}}{c}+{\sum_{i=1}^{N}}g_{ME}\left(\mathbf{r}_{0},\mathbf{r}_{i}\right)\left[\mathbf{n}_{r_{0}r_{i}}\times\mathbf{p}\left(\mathbf{r}_{i}\right)\right],\nonumber \\
 & = & \mathbf{\bar{\bar{G}}}_{MM}\left(\mathbf{r}_{0},\mathbf{r}_{0}\right)\cdot\frac{\bm{\mu}_{t}}{c}+{\sum_{i=1}^{N}}g_{ME}\left(\mathbf{r}_{0},\mathbf{r}_{i}\right)\mathbf{n}_{r_{0}r_{i}}\times\left[A\overline{\mathbf{E}}_{\mu_{t}}\right]_{i},
\end{eqnarray}
Now by using $g_{EM}\left(\mathbf{r}_{i},\mathbf{r}_{0}\right)=g_{EM}\left(\mathbf{r}_{0},\mathbf{r}_{i}\right)=-g_{ME}\left(\mathbf{r}_{0},\mathbf{r}_{i}\right)$,
the radiated power read as
\begin{eqnarray}
P_{\mathrm{rad}}^{\mathrm{ant}} & = & \frac{\omega}{2}\mathrm{Im}\left[\frac{\bm{\mu}_{t}}{c}^{*}\cdot Z_{0}\mathbf{H}_{\mathrm{local}}\left(\mathbf{r}_{0}\right)\right],\label{eq:P_rad_MD_ant-1}\\
 & = & \frac{\omega}{2}\left|\frac{\bm{\mu}_{t}}{c}\right|^{2}\mathrm{Im}\left[\mathbf{n}_{\mu}^{T}\mathbf{\bar{\bar{G}}}_{MM}\left(\mathbf{r}_{0},\mathbf{r}_{0}\right)\mathbf{n}_{\mu}\right]+\nonumber \\
 &  & +\frac{\omega}{2}\mathrm{Im}{\sum_{i=1}^{N}}g_{ME}\left(\mathbf{r}_{0},\mathbf{r}_{i}\right)\frac{\bm{\mu}_{t}}{c}^{*}\cdot\left\{ \mathbf{n}_{r_{0}r_{i}}\times\left[A\overline{\mathbf{E}}_{\mu_{t}}\right]_{i}\right\} ,
\end{eqnarray}
where $\mathrm{Im}\left[\mathbf{n}_{\mu}^{T}\mathbf{G}_{MM}\left(\mathbf{r}_{0},\mathbf{r}_{0}\right)\mathbf{n}_{\mu}\right]=\frac{1}{\epsilon_{0}\alpha_{0}}.$
Now, by using Eq.~\ref{eq:P_rad_MD_fs-1} and Eq.~\ref{eq:P_rad_MD_ant-1},
the emission rate enhancement can be obtained as
\begin{eqnarray}
\frac{P_{\mathrm{rad}}^{\mathrm{ant}}}{P_{\mathrm{rad}}^{\mathrm{fs}}} & = & 1+\frac{\epsilon_{0}\alpha_{0}}{\left|\frac{\bm{\mu}_{t}}{c}\right|}\mathrm{Im}{\sum_{i=1}^{N}}g_{ME}\left(\mathbf{r}_{0},\mathbf{r}_{i}\right)\mathbf{n}_{\mu}\cdot\left\{ \mathbf{n}_{r_{0}r_{i}}\times\left[A\overline{\mathbf{E}}_{\mu_{t}}\right]_{i}\right\} .\label{eq:PF_ED-1}
\end{eqnarray}
In the following subsections, we consider a magnetic dipole emitter
in the middle of i) an atomic dimer and ii) an atomic tetramer.

\subsection{Atomic dimer}

Let us consider a test magnetic dipole emitter $\bm{\mu}_{t}=\mu_{t}\mathbf{e}_{y}$
placed at $\mathbf{r}_{0}=\mathbf{0}$ close to an antenna consist
of two atoms with only electric dipole response (electric polarizability
$\alpha$) at position$\mathbf{r}_{u/d}=\pm l/2\mathbf{e}_{z}$ {[}see
Fig.~\ref{fig:AtomicDimerGeometry} (a){]}. Now by using Eq.~\ref{eq:PF_ED-1},
the emission rate enhancement read as

\begin{eqnarray}
\frac{P_{\mathrm{rad}}^{\mathrm{ant}}}{P_{\mathrm{rad}}^{\mathrm{fs}}} & = & 1+\frac{\epsilon_{0}\alpha_{0}}{\left|\frac{\bm{\mu}_{t}}{c}\right|}\mathrm{Im}{\sum_{i=1}^{2}}\left\{ g_{ME}\left(\mathbf{r}_{0},\mathbf{r}_{i}\right)\mathbf{n}_{\mu}\cdot\left[\mathbf{n}_{r_{0}r_{i}}\times\mathbf{p}\left(\mathbf{r}_{i}\right)\right]\right\} ,\nonumber \\
 & = & 1+\frac{\epsilon_{0}\alpha_{0}}{\left|\frac{\bm{\mu}_{t}}{c}\right|}\mathrm{Im}{\sum_{i=1}^{2}}\left\{ g_{ME}\left(\mathbf{r}_{0},\mathbf{r}_{u}\right)\mathbf{e}_{y}\cdot\left[\mathbf{n}_{r_{0}r_{u}}\times\mathbf{p}\left(\mathbf{r}_{u}\right)\right]+g_{ME}\left(\mathbf{r}_{0},\mathbf{r}_{d}\right)\mathbf{e}_{y}\cdot\left[\mathbf{n}_{r_{0}r_{d}}\times\mathbf{p}\left(\mathbf{r}_{d}\right)\right]\right\} 
\end{eqnarray}

where $\mathbf{n}_{r_{0}r_{u}}=-\mathbf{e}_{z}$, $\mathbf{n}_{r_{0}r_{d}}=\mathbf{e}_{z}$,
$\mathbf{n}_{\mu}=\mathbf{e}_{y}$ and the collective polarizability
read as

\begin{figure}
\begin{centering}
\includegraphics[width=10cm]{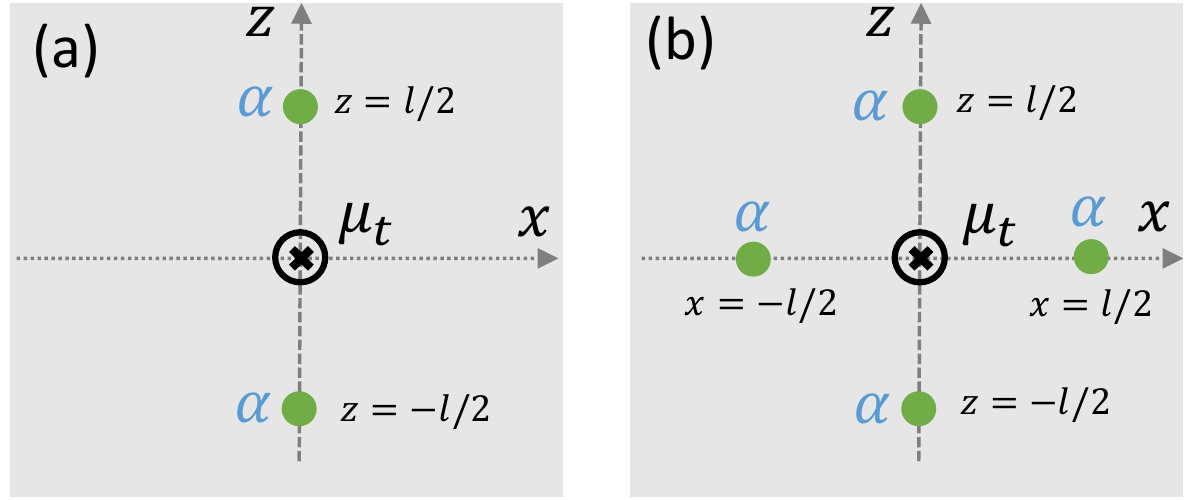}
\par\end{centering}
\caption{A test magnetic dipole emitter placed in the middle of an atomic dimer
(a) and tetramer (b) consisting of two/four identical atoms with \textit{only}
electric dipole moments, respectively. \label{fig:AtomicDimerGeometry}}
\end{figure}

\begin{equation}
A=\left[\begin{array}{cccccc}
\frac{\epsilon_{0}\alpha}{1-\epsilon_{0}^{2}\alpha^{2}G_{EE}^{xx}\left(\mathbf{r}_{u},\mathbf{r}_{d}\right)^{2}} & 0 & 0 & \frac{\alpha^{2}G_{EE}^{xx}\left(\mathbf{r}_{u},\mathbf{r}_{d}\right)}{1-\alpha^{2}G_{EE}^{xx}\left(\mathbf{r}_{u},\mathbf{r}_{d}\right)^{2}} & 0 & 0\\
0 & \frac{\epsilon_{0}\alpha}{1-\epsilon_{0}^{2}\alpha^{2}G_{EE}^{yy}\left(\mathbf{r}_{u},\mathbf{r}_{d}\right)^{2}} & 0 & 0 & \frac{\alpha^{2}G_{EE}^{yy}\left(\mathbf{r}_{u},\mathbf{r}_{d}\right)}{1-\alpha^{2}G_{EE}^{yy}\left(\mathbf{r}_{u},\mathbf{r}_{d}\right)^{2}} & 0\\
0 & 0 & \frac{\epsilon_{0}\alpha}{1-\epsilon_{0}^{2}\alpha^{2}G_{EE}^{zz}\left(\mathbf{r}_{u},\mathbf{r}_{d}\right)^{2}} & 0 & 0 & \frac{\alpha^{2}G_{EE}^{zz}\left(\mathbf{r}_{u},\mathbf{r}_{d}\right)}{1-\alpha^{2}G_{EE}^{zz}\left(\mathbf{r}_{u},\mathbf{r}_{d}\right)^{2}}\\
\frac{\alpha^{2}G_{EE}^{xx}\left(\mathbf{r}_{u},\mathbf{r}_{d}\right)}{1-\alpha^{2}G_{EE}^{xx}\left(\mathbf{r}_{u},\mathbf{r}_{d}\right)^{2}} & 0 & 0 & \frac{\epsilon_{0}\alpha}{1\epsilon_{0}^{2}\alpha^{2}G_{EE}^{xx}\left(\mathbf{r}_{u},\mathbf{r}_{d}\right)^{2}} & 0 & 0\\
0 & \frac{\alpha^{2}G_{EE}^{yy}\left(\mathbf{r}_{u},\mathbf{r}_{d}\right)}{1-\alpha^{2}G_{EE}^{yy}\left(\mathbf{r}_{u},\mathbf{r}_{d}\right)^{2}} & 0 & 0 & \frac{\epsilon_{0}\alpha}{1-\epsilon_{0}^{2}\alpha^{2}G_{EE}^{yy}\left(\mathbf{r}_{u},\mathbf{r}_{d}\right)^{2}} & 0\\
0 & 0 & \frac{\alpha^{2}G_{EE}^{zz}\left(\mathbf{r}_{u},\mathbf{r}_{d}\right)}{1-\alpha^{2}G_{EE}^{zz}\left(\mathbf{r}_{u},\mathbf{r}_{d}\right)^{2}} & 0 & 0 & \frac{\epsilon_{0}\alpha}{1-\epsilon_{0}^{2}\alpha^{2}G_{EE}^{zz}\left(\mathbf{r}_{u},\mathbf{r}_{d}\right)^{2}}
\end{array}\right].
\end{equation}

Now by using $A$ and $\mathbf{E}_{\bm{\mu}_{t}}\left(\mathbf{r}_{i}\right)\equiv g_{EM}\left(\mathbf{r}_{i},\mathbf{r}_{0}\right)\left(\mathbf{n}_{r_{i}r_{0}}\times\frac{\bm{\mu}_{t}}{c}\right)$
we obtain the dipole moments

\begin{eqnarray}
\left[\begin{array}{c}
\mathbf{p}\left(\mathbf{r}_{u}\right)\\
\mathbf{p}\left(\mathbf{r}_{d}\right)
\end{array}\right] & = & A\left[\begin{array}{c}
g_{EM}\left(\mathbf{r}_{u},\mathbf{r}_{0}\right)\left(\mathbf{n}_{r_{u}r_{0}}\times\frac{\bm{\mu}_{t}}{c}\right)\\
g_{EM}\left(\mathbf{r}_{d},\mathbf{r}_{0}\right)\left(\mathbf{n}_{r_{d}r_{0}}\times\frac{\bm{\mu}_{t}}{c}\right)
\end{array}\right],\\
 & = & \frac{\epsilon_{0}\alpha\left|\frac{\bm{\mu}_{t}}{c}\right|}{1+\epsilon_{0}\alpha G_{EE}^{xx}\left(\mathbf{r}_{u},\mathbf{r}_{d}\right)}\left[\begin{array}{c}
-g_{EM}\left(\mathbf{r}_{u},\mathbf{r}_{0}\right)\\
0\\
0\\
g_{EM}\left(\mathbf{r}_{d},\mathbf{r}_{0}\right)\\
0\\
0
\end{array}\right],
\end{eqnarray}
we used
\begin{equation}
\left[\begin{array}{c}
\mathbf{n}_{r_{u}r_{0}}\times\frac{\bm{\mu}_{t}}{c}\\
\mathbf{n}_{r_{d}r_{0}}\times\frac{\bm{\mu}_{t}}{c}
\end{array}\right]=\left[\begin{array}{cccccc}
-1 & 0 & 0 & 1 & 0 & 0\end{array}\right]^{T},
\end{equation}
and we get 
\begin{eqnarray}
\left[\begin{array}{c}
\mathbf{n}_{r_{0}r_{u}}\times\mathbf{p}\left(\mathbf{r}_{u}\right)\\
\mathbf{n}_{r_{0}r_{d}}\times\mathbf{p}\left(\mathbf{r}_{d}\right)
\end{array}\right] & = & \frac{\epsilon_{0}\alpha\left|\frac{\bm{\mu}_{t}}{c}\right|}{1+\epsilon_{0}\alpha G_{EE}^{xx}\left(\mathbf{r}_{u},\mathbf{r}_{d}\right)}\left[\begin{array}{cccccc}
0 & g_{EM}\left(\mathbf{r}_{u},\mathbf{r}_{0}\right) & 0 & 0 & g_{EM}\left(\mathbf{r}_{d},\mathbf{r}_{0}\right) & 0\end{array}\right]^{T}.
\end{eqnarray}
Finally, we obtain
\begin{eqnarray}
\frac{P_{\mathrm{rad}}^{\mathrm{ant}}}{P_{\mathrm{rad}}^{\mathrm{fs}}} & = & 1+\frac{\epsilon_{0}\alpha_{0}}{\left|\frac{\bm{\mu}_{t}}{c}\right|}\mathrm{Im}{\sum_{i=1}^{2}}\left\{ g_{ME}\left(\mathbf{r}_{0},\mathbf{r}_{u}\right)\mathbf{e}_{y}\cdot\left[\mathbf{n}_{r_{0}r_{u}}\times\mathbf{p}\left(\mathbf{r}_{u}\right)\right]+g_{ME}\left(\mathbf{r}_{0},\mathbf{r}_{d}\right)\mathbf{e}_{y}\cdot\left[\mathbf{n}_{r_{0}r_{d}}\times\mathbf{p}\left(\mathbf{r}_{d}\right)\right]\right\} ,\nonumber \\
 & = & 1+\epsilon_{0}\alpha_{0}\mathrm{Im}\left[\frac{g_{EM}\left(\mathbf{r}_{u},\mathbf{r}_{0}\right)g_{ME}\left(\mathbf{r}_{0},\mathbf{r}_{u}\right)+g_{EM}\left(\mathbf{r}_{d},\mathbf{r}_{0}\right)g_{ME}\left(\mathbf{r}_{0},\mathbf{r}_{d}\right)}{\frac{1}{\epsilon_{0}\alpha}+G_{EE}^{xx}\left(\mathbf{r}_{u},\mathbf{r}_{d}\right)}\right],\nonumber \\
 & = & 1-2\epsilon_{0}^{2}\alpha_{0}\alpha\mathrm{Im}\left[\frac{g_{EM}^{2}\left(\mathbf{r}_{0},\mathbf{r}_{u}\right)}{1+\epsilon_{0}\alpha G_{EE}^{xx}\left(\mathbf{r}_{u},\mathbf{r}_{d}\right)}\right],\nonumber \\
 & = & 1-2\epsilon_{0}^{2}\alpha_{0}\alpha\mathrm{Im}\left[\frac{g_{EM}^{2}\left(\mathbf{r}_{0},\mathbf{r}_{u}\right)}{D_{+}}\right],\label{Decay_dimer}
\end{eqnarray}
where $D_{+}\equiv1+\epsilon_{0}\alpha G_{EE}^{xx}\left(\mathbf{r}_{u},\mathbf{r}_{d}\right)$
and $g_{EM}\left(\mathbf{r}_{u},\mathbf{r}_{0}\right)=-g_{ME}\left(\mathbf{r}_{0},\mathbf{r}_{u}\right)$. We used Eq.~\ref{Decay_dimer} to calculate the decay rate in Fig.~2 of the main manuscript. The Green functions for the atomic dimer (see Fig.~\ref{fig:AtomicDimerGeometry})
read as
\begin{eqnarray}
G_{EE}^{xx}\left(\mathbf{r}_{u},\mathbf{r}_{d}\right) & = & \frac{3}{2\epsilon_{0}\alpha_{0}}e^{i\zeta}\left(\frac{1}{\zeta}-\frac{1}{\zeta^{3}}+\frac{i}{\zeta^{2}}\right),\,\,\,\zeta=k\left|\mathbf{r}_{u}-\mathbf{r}_{d}\right|=kl,\\
g_{ME}\left(\mathbf{r}_{0},\mathbf{r}_{u}\right) & = & g_{ME}\left(\mathbf{r}_{0},\mathbf{r}_{l}\right)=\frac{3}{2\epsilon_{0}\alpha_{0}}e^{i\zeta}\left(\frac{1}{\zeta}+\frac{i}{\zeta^{2}}\right)\,\,\,\,\zeta=k\left|\mathbf{r}_{0}-\mathbf{r}_{u}\right|=k\left|\mathbf{r}_{0}-\mathbf{r}_{d}\right|=kl/2.\nonumber 
\end{eqnarray}
Finally, we employ the link between the quantum (i.e.
decay rate) and classical formalisms (i.e. radiated power)~\cite{novotny2012},
i.e.
\begin{equation}
\boxed{\frac{\Gamma_{\mathrm{\rm ant}}}{\Gamma_{0}}=\frac{P_{\mathrm{rad}}^{\mathrm{ant}}}{P_{\mathrm{rad}}^{\mathrm{fs}}}= 1-2\epsilon_{0}^{2}\alpha_{0}\alpha\mathrm{Im}\left[\frac{g_{EM}^{2}\left(\mathbf{r}_{0},\mathbf{r}_{u}\right)}{D_{+}}\right]\label{eq:Decay_Rad_dimer}}
\end{equation}
\subsection{Atomic tetramer}
Let us consider a magnetic dipole emitter $\bm{\mu}_{t}=\mu_{t}\mathbf{e}_{y}$
placed at $\mathbf{r}_{0}=\mathbf{0}$ close to four quantum antennas
(i.e. electric dipole moments) with polarizability $\alpha$ at position
$\mathbf{r}_{1,2}=\mp l/2\mathbf{e}_{z}$, $\mathbf{r}_{3,4}=\pm l/2\mathbf{e}_{x}$
{[}see Fig.~\ref{fig:Geometry_dimer} (b){]}. Now by using Eq.~\ref{eq:PF_ED-1},
the emission rate enhancement read as
\begin{eqnarray}
\frac{P_{\mathrm{rad}}^{\mathrm{ant}}}{P_{\mathrm{rad}}^{\mathrm{fs}}} & = & 1+\frac{\epsilon_{0}\alpha_{0}}{\left|\frac{\bm{\mu}_{t}}{c}\right|}\mathrm{Im}\sum_{i=1}^{4}\left\{ g_{ME}\left(\mathbf{r}_{0},\mathbf{r}_{i}\right)\mathbf{e}_{y}\cdot\left[\mathbf{n}_{r_{0}r_{i}}\times\mathbf{p}\left(\mathbf{r}_{i}\right)\right]\right\} ,\\
 & = & 1+\frac{\epsilon_{0}\alpha_{0}}{\left|\frac{\bm{\mu}_{t}}{c}\right|}\mathrm{Im}\sum_{i=1}^{4}g_{ME}\left(\mathbf{r}_{0},\mathbf{r}_{i}\right)\mathbf{n}_{\mu}\cdot\left\{ \mathbf{n}_{r_{0}r_{i}}\times\left[A\overline{\mathbf{E}}_{\mu_{t}}\right]_{i}\right\} 
\end{eqnarray}
where $\mathbf{n}_{\mu}=\mathbf{e}_{y}$ and the normal unit vectors
read as
\begin{equation}
\mathbf{n}_{r_{0}r_{1}}=\left[\begin{array}{c}
0\\
0\\
1
\end{array}\right],\,\,\,\,\mathbf{n}_{r_{0}r_{2}}=\left[\begin{array}{c}
0\\
0\\
-1
\end{array}\right],\,\,\,\,\mathbf{n}_{r_{0}r_{3}}=\left[\begin{array}{c}
-1\\
0\\
0
\end{array}\right],\,\,\,\,\mathbf{n}_{r_{0}r_{4}}=\left[\begin{array}{c}
1\\
0\\
0
\end{array}\right],
\end{equation}
$A$ matrix read as
\begin{equation}
A=\left[\begin{array}{cccc}
\mathbf{\bar{\bar{I}}}/\alpha & \mathbf{\bar{\bar{G}}}_{EE}\left(\mathbf{r}_{1},\mathbf{r}_{2}\right) & \mathbf{\bar{\bar{G}}}_{EE}\left(\mathbf{r}_{1},\mathbf{r}_{3}\right) & \mathbf{\bar{\bar{G}}}_{EE}\left(\mathbf{r}_{1},\mathbf{r}_{4}\right)\\
\mathbf{\bar{\bar{G}}}_{EE}\left(\mathbf{r}_{1},\mathbf{r}_{2}\right) & \mathbf{\bar{\bar{I}}}/\alpha & \mathbf{\bar{\bar{G}}}_{EE}\left(\mathbf{r}_{2},\mathbf{r}_{3}\right) & \mathbf{\bar{\bar{G}}}_{EE}\left(\mathbf{r}_{2},\mathbf{r}_{4}\right)\\
\mathbf{\bar{\bar{G}}}_{EE}\left(\mathbf{r}_{1},\mathbf{r}_{3}\right) & \mathbf{\bar{\bar{G}}}_{EE}\left(\mathbf{r}_{2},\mathbf{r}_{3}\right) & \mathbf{\bar{\bar{I}}}/\alpha & \mathbf{\bar{\bar{G}}}_{EE}\left(\mathbf{r}_{3},\mathbf{r}_{4}\right)\\
\mathbf{\bar{\bar{G}}}_{EE}\left(\mathbf{r}_{1},\mathbf{r}_{4}\right) & \mathbf{\bar{\bar{G}}}_{EE}\left(\mathbf{r}_{2},\mathbf{r}_{4}\right) & \mathbf{\bar{\bar{G}}}_{EE}\left(\mathbf{r}_{3},\mathbf{r}_{4}\right) & \mathbf{\bar{\bar{I}}}/\alpha
\end{array}\right],
\end{equation}
where ${\bf \bar{\bar{I}}}$ is the identity dyadic

\begin{equation}
\mathbf{\bar{\bar{G}}}_{EE}\left(\mathbf{r}_{1},\mathbf{r}_{2}\right)=\frac{3}{2\alpha_{0}\epsilon_{0}}e^{ikl}\left[\begin{array}{ccc}
g_{1}\left(kl\right) & 0 & 0\\
0 & g_{1}\left(kl\right) & 0\\
0 & 0 & g_{1}\left(kl\right)+g_{2}\left(kl\right)
\end{array}\right],
\end{equation}

\begin{equation}
\mathbf{\bar{\bar{G}}}_{EE}\left(\mathbf{r}_{3},\mathbf{r}_{4}\right)=\frac{3}{2\alpha_{0}\epsilon_{0}}e^{ikl}\left[\begin{array}{ccc}
g_{1}\left(kl\right)+g_{2}\left(kl\right) & 0 & 0\\
0 & g_{1}\left(kl\right) & 0\\
0 & 0 & g_{1}\left(kl\right)
\end{array}\right],
\end{equation}

\begin{equation}
\mathbf{\bar{\bar{G}}}_{EE}\left(\mathbf{r}_{1},\mathbf{r}_{3}\right)=\mathbf{\bar{\bar{G}}}_{EE}\left(\mathbf{r}_{2},\mathbf{r}_{4}\right)=\frac{3}{2\alpha_{0}\epsilon_{0}}e^{i\frac{kl}{\sqrt{2}}}\left[\begin{array}{ccc}
g_{1}\left(\frac{kl}{\sqrt{2}}\right) & 0 & \frac{g_{2}\left(\frac{kl}{\sqrt{2}}\right)}{2}\\
0 & g_{1}\left(\frac{kl}{\sqrt{2}}\right) & 0\\
\frac{g_{2}\left(\frac{kl}{\sqrt{2}}\right)}{2} & 0 & g_{1}\left(\frac{kl}{\sqrt{2}}\right)
\end{array}\right],
\end{equation}

\begin{equation}
\mathbf{\bar{\bar{G}}}_{EE}\left(\mathbf{r}_{1},\mathbf{r}_{4}\right)=\mathbf{\bar{\bar{G}}}_{EE}\left(\mathbf{r}_{2},\mathbf{r}_{3}\right)=\frac{3}{2\alpha_{0}\epsilon_{0}}e^{i\frac{kl}{\sqrt{2}}}\left[\begin{array}{ccc}
g_{1}\left(\frac{kl}{\sqrt{2}}\right) & 0 & -\frac{g_{2}\left(\frac{kl}{\sqrt{2}}\right)}{2}\\
0 & g_{1}\left(\frac{kl}{\sqrt{2}}\right) & 0\\
-\frac{g_{2}\left(\frac{kl}{\sqrt{2}}\right)}{2} & 0 & g_{1}\left(\frac{kl}{\sqrt{2}}\right)
\end{array}\right],
\end{equation}
we used the Green function definition, i.e.

\begin{eqnarray}
G_{EE}^{\alpha\beta}\left(\zeta=k\left|\mathbf{r}-\mathbf{r}^{\prime}\right|\right) & = & \frac{3}{2\alpha_{0}\epsilon_{0}}e^{iu}\left[g_{1}\left(\zeta\right)\delta_{\alpha\beta}+g_{2}\left(\zeta\right)\frac{\zeta_{\alpha}\zeta_{\beta}}{\zeta^{2}}\right],\nonumber \\
g_{1}\left(\zeta\right) & = & \left(\frac{1}{\zeta}-\frac{1}{\zeta^{3}}+\frac{i}{\zeta^{2}}\right),\nonumber \\
g_{2}\left(\zeta\right) & = & \left(-\frac{1}{\zeta}+\frac{3}{\zeta^{3}}-\frac{3i}{\zeta^{2}}\right).
\end{eqnarray}

Thus the induced dipole moment read as
\[
\left[\begin{array}{c}
\mathbf{p}\left(\mathbf{r}_{1}\right)\\
\mathbf{p}\left(\mathbf{r}_{2}\right)\\
\mathbf{p}\left(\mathbf{r}_{3}\right)\\
\mathbf{p}\left(\mathbf{r}_{4}\right)
\end{array}\right]=A\left[\begin{array}{c}
g_{EM}\left(\mathbf{r}_{1},\mathbf{r}_{0}\right)\left(\mathbf{n}_{r_{1}r_{0}}\times\frac{\bm{\mu}_{t}}{c}\right)\\
g_{EM}\left(\mathbf{r}_{2},\mathbf{r}_{0}\right)\left(\mathbf{n}_{r_{2}r_{0}}\times\frac{\bm{\mu}_{t}}{c}\right)\\
g_{EM}\left(\mathbf{r}_{3},\mathbf{r}_{0}\right)\left(\mathbf{n}_{r_{3}r_{0}}\times\frac{\bm{\mu}_{t}}{c}\right)\\
g_{EM}\left(\mathbf{r}_{4},\mathbf{r}_{0}\right)\left(\mathbf{n}_{r_{4}r_{0}}\times\frac{\bm{\mu}_{t}}{c}\right)
\end{array}\right],
\]
and we get
\begin{eqnarray}
\mathbf{p}\left(\mathbf{r}_{1}\right) & = & \frac{\epsilon_{0}\alpha\left|\frac{\bm{\mu}_{t}}{c}\right|g_{EM}\left(\mathbf{r}_{1},\mathbf{r}_{0}\right)}{1+\frac{3\alpha}{2\alpha_{0}}g_{1}\left(kl\right)e^{ikl}-\frac{3\alpha}{2\alpha_{0}}g_{2}\left(\frac{kl}{\sqrt{2}}\right)e^{i\frac{kl}{\sqrt{2}}}}\mathbf{e}_{x},\nonumber \\
\mathbf{p}\left(\mathbf{r}_{2}\right) & = & -\frac{\epsilon_{0}\alpha\left|\frac{\bm{\mu}_{t}}{c}\right|g_{EM}\left(\mathbf{r}_{2},\mathbf{r}_{0}\right)}{1+\frac{3\alpha}{2\alpha_{0}}g_{1}\left(kl\right)e^{ikl}-\frac{3\alpha}{2\alpha_{0}}g_{2}\left(\frac{kl}{\sqrt{2}}\right)e^{i\frac{kl}{\sqrt{2}}}}\mathbf{e}_{x},\nonumber \\
\mathbf{p}\left(\mathbf{r}_{3}\right) & = & \frac{\epsilon_{0}\alpha\left|\frac{\bm{\mu}_{t}}{c}\right|g_{EM}\left(\mathbf{r}_{3},\mathbf{r}_{0}\right)}{1+\frac{3\alpha}{2\alpha_{0}}g_{1}\left(kl\right)e^{ikl}-\frac{3\alpha}{2\alpha_{0}}g_{2}\left(\frac{kl}{\sqrt{2}}\right)e^{i\frac{kl}{\sqrt{2}}}}\mathbf{e}_{z},\nonumber \\
\mathbf{p}\left(\mathbf{r}_{4}\right) & = & -\frac{\epsilon_{0}\alpha\left|\frac{\bm{\mu}_{t}}{c}\right|g_{EM}\left(\mathbf{r}_{4},\mathbf{r}_{0}\right)}{1+\frac{3\alpha}{2\alpha_{0}}g_{1}\left(kl\right)e^{ikl}-\frac{3\alpha}{2\alpha_{0}}g_{2}\left(\frac{kl}{\sqrt{2}}\right)e^{i\frac{kl}{\sqrt{2}}}}\mathbf{e}_{z},
\end{eqnarray}
we get
\begin{eqnarray}
\sum_{i=1}^{4}
\mathbf{e}_{y}\cdot\left[\mathbf{n}_{r_{0}r_{i}}\times\mathbf{p}\left(\mathbf{r}_{i}\right)\right] & = & \epsilon_{0}\alpha\left|\frac{\bm{\mu}_{t}}{c}\right|\frac{g_{EM}\left(\mathbf{r}_{1},\mathbf{r}_{0}\right)+g_{EM}\left(\mathbf{r}_{2},\mathbf{r}_{0}\right)+g_{EM}\left(\mathbf{r}_{3},\mathbf{r}_{0}\right)+g_{EM}\left(\mathbf{r}_{4},\mathbf{r}_{0}\right)}{1+\frac{3\alpha}{2\alpha_{0}}g_{1}\left(kl\right)e^{ikl}-\frac{3\alpha}{2\alpha_{0}}e^{i\frac{kl}{\sqrt{2}}}g_{2}\left(\frac{kl}{\sqrt{2}}\right)}.
\end{eqnarray}
Using $g_{EM}\left(\mathbf{r}_{1},\mathbf{r}_{0}\right)=g_{EM}\left(\mathbf{r}_{2},\mathbf{r}_{0}\right)=g_{EM}\left(\mathbf{r}_{3},\mathbf{r}_{0}\right)=g_{EM}\left(\mathbf{r}_{4},\mathbf{r}_{0}\right)$,
we obtain the emission rate enhancement
\begin{eqnarray}
\frac{P_{\mathrm{rad}}^{\mathrm{ant}}}{P_{\mathrm{rad}}^{\mathrm{fs}}} & = & 1+\frac{\epsilon_{0}\alpha_{0}}{\left|\frac{\bm{\mu}_{t}}{c}\right|}\mathrm{Im}\sum_{i=1}^{4}\left\{ g_{ME}\left(\mathbf{r}_{0},\mathbf{r}_{i}\right)\mathbf{e}_{y}\cdot\left[\mathbf{n}_{r_{0}r_{i}}\times\mathbf{p}\left(\mathbf{r}_{i}\right)\right]\right\} ,\\
 & = & 1-4\epsilon_{0}\alpha_{0}\mathrm{Im}\left[\frac{\epsilon_{0}\alpha g_{EM}^{2}\left(\mathbf{r}_{0},\mathbf{r}_{1}\right)}{1+\frac{3\alpha}{2\alpha_{0}}g_{1}\left(kl\right)e^{ikl}-\frac{3\alpha}{2\alpha_{0}}g_{2}\left(\frac{kl}{\sqrt{2}}\right)e^{i\frac{kl}{\sqrt{2}}}}\right].
\end{eqnarray}
where $g_{ME}\left(\zeta=k\left|\mathbf{r}_{0}-\mathbf{r}_{1}\right|\right)=-g_{EM}\left(\zeta=k\left|\mathbf{r}_{0}-\mathbf{r}_{1}\right|\right)=\frac{3}{2\alpha_{0}\epsilon_{0}}e^{i\zeta}\left(\frac{1}{\zeta}-\frac{1}{i\zeta^{2}}\right)$,
and $\zeta=k\left|\mathbf{r}_{0}-\mathbf{r}_{1}\right|=\frac{kl}{2}$. Finally, we employ the link between the quantum and classical formalisms~\cite{novotny2012},
i.e.
\begin{equation}
\boxed{\frac{\Gamma_{\mathrm{\rm ant}}}{\Gamma_{0}}=1-4\epsilon_{0}\alpha_{0}\mathrm{Im}\left[\frac{\epsilon_{0}\alpha g_{EM}^{2}\left(\mathbf{r}_{0},\mathbf{r}_{1}\right)}{1+\frac{3\alpha}{2\alpha_{0}}g_{1}\left(kl\right)e^{ikl}-\frac{3\alpha}{2\alpha_{0}}g_{2}\left(\frac{kl}{\sqrt{2}}\right)e^{i\frac{kl}{\sqrt{2}}}}\right]\label{eq:Decay_Rad_tetramer}}
\end{equation}
\\
\end{widetext}
%
\end{document}